\documentclass[aps,twocolumn,floatfix]{revtex4}

\usepackage{amsmath}
\usepackage{latexsym}
\usepackage{float}
\usepackage{amssymb}
\usepackage{graphicx}
\usepackage{epsfig}
\usepackage{color}

\definecolor{red}{rgb}{0.0,0.0,0.0}

\begin{document}

\title{Monte Carlo Simulations of Lattice Models for Single Polymer Systems}

\author{Hsiao-Ping Hsu} 
\affiliation{Max-Planck-Institut f\"ur Polymerforschung, Ackermannweg 10, D-55128 Mainz, Germany}
\email{hsu@mpip-mainz.mpg.de}

\begin{abstract}
Single linear polymer chains in dilute solutions under good solvent conditions
are studied by Monte Carlo simulations with the pruned-enriched Rosenbluth
method up to the chain length $N \sim {\cal O}(10^4)$.
Based on the standard simple cubic lattice model (SCLM) with fixed bond length
and the bond fluctuation model (BFM) with bond lengths in a range between
$2$ and $\sqrt{10}$,
we investigate the conformations of polymer chains described by 
self-avoiding walks (SAWs) on the simple cubic lattice, and by random 
walks (RWs) and non-reversible random walks (NRRWs) in the absence
of excluded volume (EV) interactions.
In addition to flexible chains, we also extend our study to semiflexible
chains for different stiffness controlled by a bending potential.
The persistence lengths of chains extracted from the orientational
correlations are estimated for all cases. We show that chains based on
the BFM are more flexible than those based on the SCLM for a fixed bending energy.
The microscopic differences between these two lattice models
are discussed and the theoretical predictions of scaling laws given 
in the literature are checked and verified. 
Our simulations clarify that a different mapping
ratio between the coarse-grained models and the atomistically realistic 
description of polymers is required in a coarse-graining approach
due to the different crossovers to the asymptotic behavior.

\end{abstract}

\maketitle

\section{Introduction}

In the theoretical study of polymer physics~\cite{Flory1969,deGennes1979}, 
computer simulations provide a powerful method
to mimic the behavior of polymers covering the range from atomic to coarse-grained scales
depending on the problems one is interested in~\cite{Binder1995, Binder2008}.
The generic scaling properties of single linear and branched polymers in the bulk or 
confinement under various solvent conditions have been described quite well by 
simple coarse-grained lattice models (i.e., random
walks (RWs), non-reversible random walks (NRRWs), self-avoiding random walks (SAWs), or 
interacting self-avoiding random walks (ISAWs) on a regular lattice,
regarding the interactions between non-bonded monomers).
As an alternative one can use coarse-grained models in the continuum, such as
a bead-spring model (BSM)
(where all beads interact with a truncated and shifted Lennard-Jones (LJ) potential while the
bonded interactions are captured by a finitely extensible nonlinear elastic (FENE) potential) using
Monte Carlo and molecular dynamics simulations~\cite{Binder1995}. On the one hand, however, 
as the size and complexity of a system increases, detailed information at the atomic scale may 
be lost when employing low resolution coarse-graining representations. On the other hand, 
the cost of computing time may be too high if the system is described at high resolution. 
Therefore, more scientific effort has been devoted to developing
an appropriate coarse-grained model which 
can reproduce the global thermodynamic properties and the local mechanical and chemical properties 
such as the intermolecular forces between polymer 
chains~\cite{Murat1998, Plathe2002, Harman2006,Harman2007, Gujrati2010, Vettorel2010,
Zhang2013}. 
While these models are already known since a long time, the present work is the 
first study presenting precise data on conformational properties of these 
model, when a bond angle potential is included.

In this paper we deal with linear polymer chains in dilute solutions under
good solvent conditions, and describe them by lattice models on the simple cubic lattice. 
Although coarse-grained lattice models neglect the chemical detail of a specific
polymer chain and only keep chain connectivity (topology) and excluded volume,
the universal behavior of polymers still remains the same in the thermodynamic
limit (as the chain length $N\rightarrow \infty$)~\cite{deGennes1979}, 
Two coarse-grained lattice models, the standard simple cubic lattice model
(SCLM) and the bond fluctuation model 
(BFM)~\cite{Binder1995,Carmesin1988, Deutsch1991, Paul1991, Mueller2005}, 
are considered for our simulations.
{The SCLM is often used for the test of new simulation 
algorithms, and the verification of theoretically predicted scaling laws 
due to its simplicity and computational efficiency.}
The BFM has the advantages that the computational efficiency of lattice 
models is kept and the behavior of polymers in a continuum space can be 
described approximately. The model thus introduces some local conformational 
flexibility while
retaining the computational efficiency of lattice models for implementing
excluded volume interactions by enforcing a single occupation of each 
lattice vertex.

{red}{The excluded volume effect plays an essential role
in any real polymer chain, while in a dilute solution under a theta
solvent condition, or in concentrated polymer solutions such as
melts, and glasses, the real polymer chain behaves like an ideal chain.
The excluded volume constraint can easily be incorporated in 
the lattice models by simply forbidding any two effective monomers
occupying the same lattice site (cell). Tries et al. have successfully
mapped linear polyethylene in the melt onto the BFM and their results
are in good agreement with experimental viscosimetric results 
quantitatively without adjusting any extra parameters~\cite{Mueller2005,Tries1997}.
Varying the backbone length and side chain length of the bottle-brush
polymer based on the BFM, a direct comparison of the structure factors 
between the experimental data for the synthetic bottle-brush polymer 
consisting of hydroxyethyl methacrylate (PMMA) as the backbone polymer
and flexible poly(n-butyl acrylate) (PnBA) as side chains
in a good solvent (toluene) and the Monte Carlo results
is given~\cite{Hsu2009}. Furthermore, the lattice models have also been
used widely to investigate the conformational properties of 
protein-folding~\cite{Lau1989}
and DNA in chromosomes~\cite{Vettorel2009a,Vettorel2009b,Halverson2014} in biopolymers. 
For alkane-like chains the angles between subsequent effective bonds
are not continuously distributed, but only discrete angles are allowed.
Therefore, the lattice model such as the SCLM where only the discrete
angles $0^o$ and $90^o$ are allowed is an ideal model for studying 
alkane-like chains of different stiffnesses.}
 
{A direct comparison of simulation results 
between the lattice models and the off-lattice BSM
is also possible, e.g. linear polymers~\cite{Wittmer2007} and
ring polymers~\cite{Halverson2013} in a melt,
adsorption of multi-block and random copolymers~\cite{Bhattacharya2008},
semiflexible chains under a good solvent 
condition~\cite{Huang2014c},
the crossover from semiflexible polymer brushes towards semiflexible
mushrooms as the grafting density decreases~\cite{Egorov2014}.
Results from these two coarse-grained models are qualitatively the same. Namely,
they both show the same scaling behavior, but the amplitudes and
the critical or the crossover points can vary depending on the underlying
models.
However, our main motivation was to understand 
the microscopic differences between the two lattice models, SCLM 
and BFM, for describing the conformations of polymers and 
to provide this information for the further development of
a multi-scale approach for studying polymers of complex topology and
polymer solutions at high
concentration based on the lattice models. Therefore, we only focus on
the two coarse-grained lattice models, SCLM and BFM, here.}
In the mapping between atomistic and coarse-grained models, it turns out
that a bond angle potential also needs to be included in the
coarse-grained models, as is done here.

The outline of the paper is as follows: Sec.~\ref{model}
describes the models and the simulation technique.
Sec.~\ref{single} and Sec.~\ref{semi} 
review the properties of flexible chains and semiflexible chains,
respectively. Polymer chains described by SAWs, and by RWs and NRRWs
in the absence of excluded volume effects are studied and compared 
with theoretical predictions. 
Finally our conclusions are summarized in Sec.~\ref{conclusion}.

\section{Models and simulation methods}
\label{model}
The basic characteristics of linear polymer chains depend
on the solvent conditions. Under good solvent conditions
the repulsive interactions (the excluded volume effect) and
entropic effects dominate the conformation, and the polymer
chain tends to swell to a random coil.
In the thermodynamic limit, namely the chain length
$N \rightarrow \infty$, the partition function scales as
\begin{equation}
      Z_N \sim \mu_\infty^{-N} N^{\gamma_d-1} \sim q_{\rm eff}^N N^{\gamma_d-1}
\label{eq-ZN}
\end{equation}
where $\mu_\infty$ is the critical fugacity per monomer, $q_{\rm eff}$ is
the effective coordination number,
and $\gamma$ is the entropic exponent related to the topology.
In two dimensions~\cite{deGennes1979} $\gamma=43/32$,
while the best estimate~\cite{Hsu2004} for $d=3$ is 
$\gamma=1.1573(2)$. For the standard self-avoiding
walks on the simple cubic lattice~\cite{Grassberger2005}
in $d=3$ one has $\mu_\infty=0.21349098(5)$
and the corresponding effective coordination number
$q_{\rm eff}=1/\mu_\infty=4.6840386(11)$.
The conformations of polymer chains characterized by
the mean square end-to-end distance,
$\langle R_e^2 \rangle$, and the mean square 
gyration radius, $\langle R_g^2 \rangle$,
scale as~\cite{Li1995,Nickel1991}:
\begin{equation}
   \langle R_e^2 \rangle/ \ell_b^2 
 = A_e N^{2\nu}[1+{\cal O}(N^{-\Delta})] \, ,
\label{eq-Re}
\end{equation}
and
\begin{equation}
    \langle R_g^2 \rangle /  \ell_b^2 
 = A_g N^{2\nu}[1+{\cal O}(N^{-\Delta})] 
\label{eq-Rg}
\end{equation}
where $\nu$ is the Flory exponent, $\Delta$ is the leading correction
to the scaling exponent, $A_e$ and $A_g$ are non-universal constants,
and $\ell_b^2$ is the mean square bond length.
The quantities $\nu$, $\Delta$, and the ratio $A_e/A_g$ are 
universal~\cite{Privman1991}, while the quantities, $A_e$, $A_g$, 
$\ell_b$, 
and $q_{\rm eff}$, 
depend on the microscopic realization. In $d=2$ one has 
$\nu_2=3/4$, while in $d=3$ the most accurate estimate of
the Flory exponent~\cite{Clisby2010} $\nu=0.587597(7)$.
We use $\nu=0.5876$ for our data analysis in this paper.

\begin{figure*}[t]
\begin{center}
(a)\includegraphics[scale=0.29,angle=270]{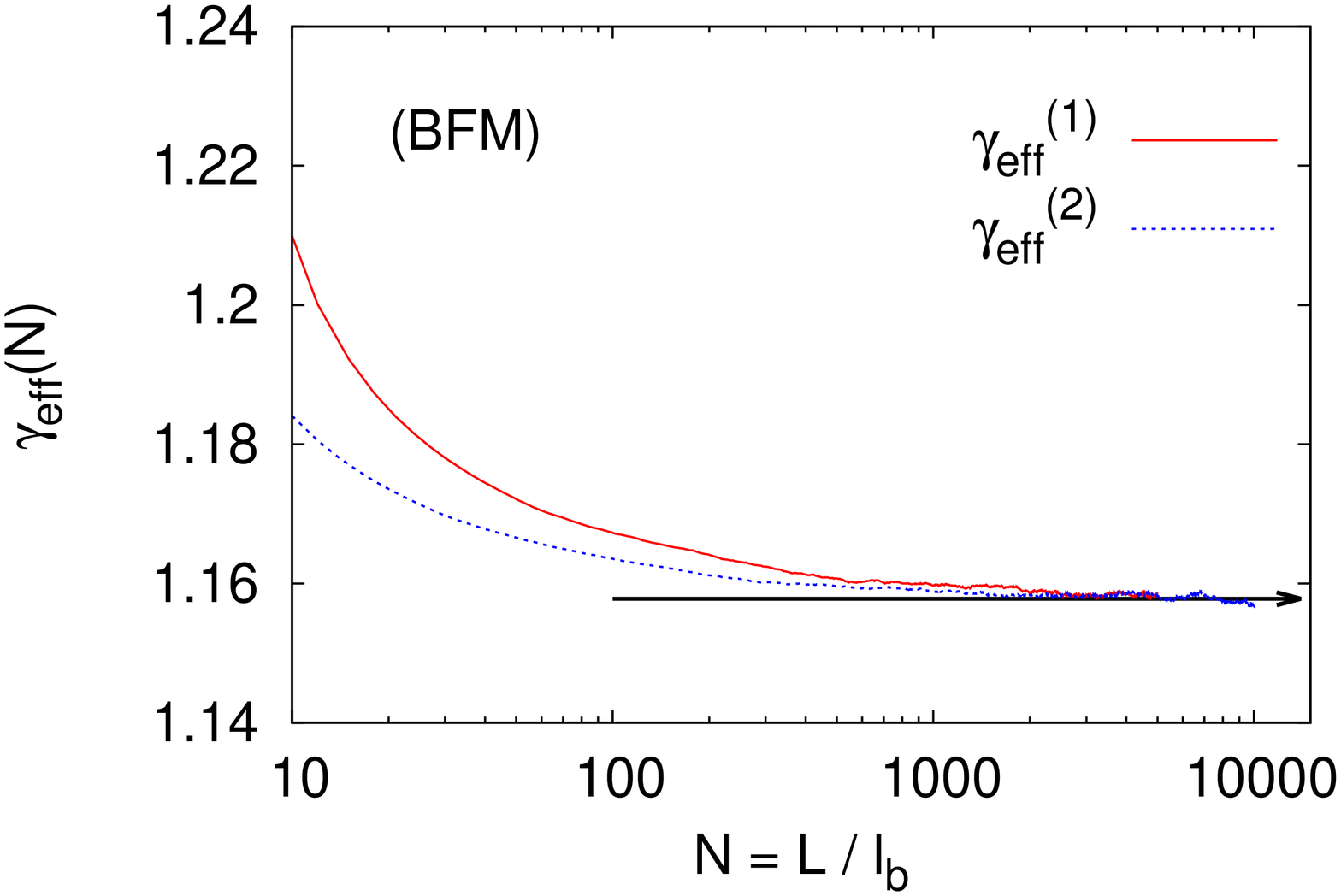} \hspace{0.4cm}
(b)\includegraphics[scale=0.29,angle=270]{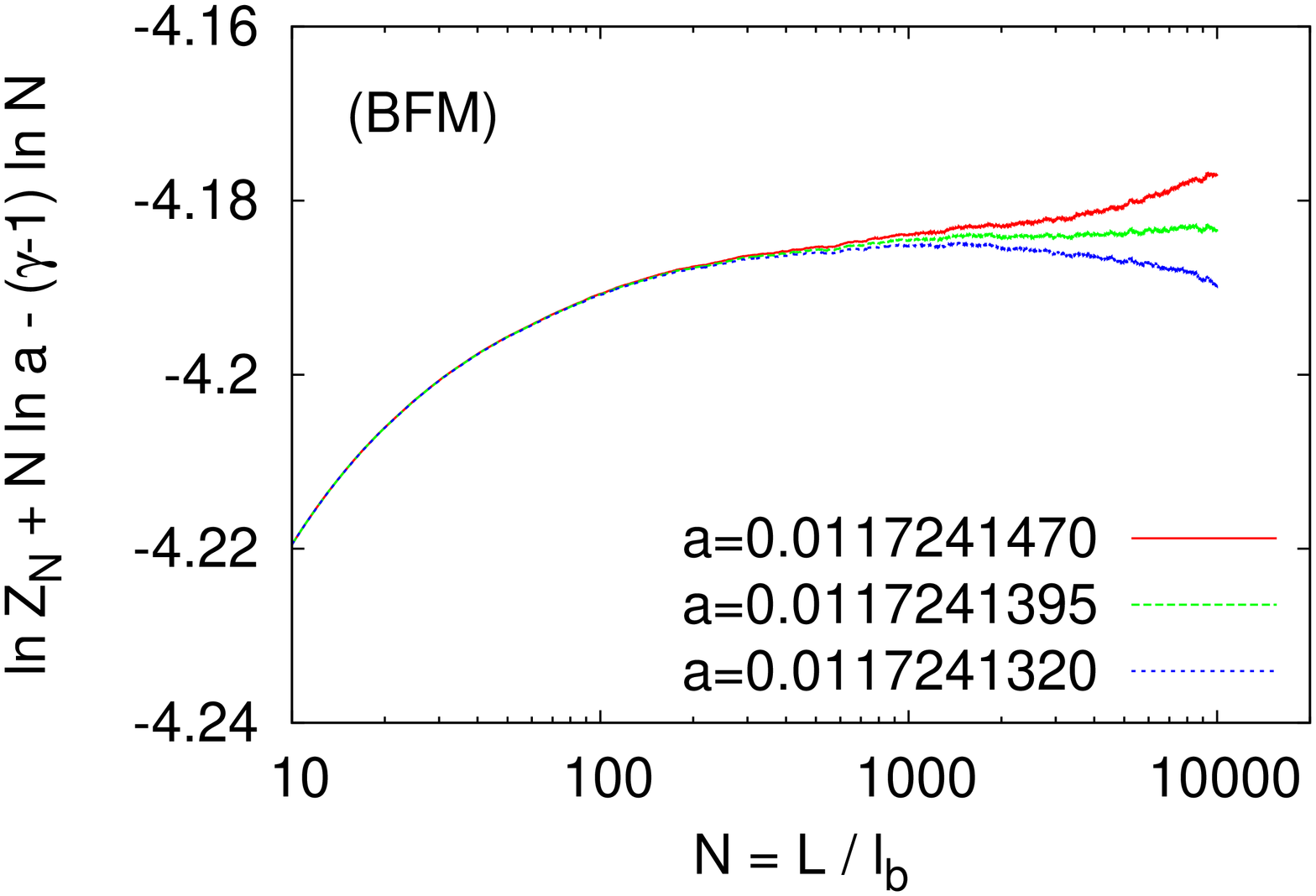} \hspace{0.4cm}
\caption{(a) Effective exponents $\gamma_{\rm eff}^{(1)}$ and
$\gamma_{\rm eff}^{(2)}$ [computed from Eqs.~(\ref{eq-gamma1}) and (\ref{eq-gamma2}) ]
plotted versus $N$ on a semi-log scale.
$\gamma=\lim_{N \rightarrow \infty} \gamma_{\rm eff}^{(1)}(N) = 
\lim_{N \rightarrow \infty} \gamma_{\rm eff}^{(2)}(N)=1.1578(6)$.
(b) $\ln Z_N + N \ln a -(\gamma-1) \ln N$ with $\gamma=1.1578$
determined from (a) plotted versus $N$ on a semi-log scale.
The best estimate of fugacity $\mu=a=0.0117241395(75)$ is determined by the
horizontal curve.}
\label{fig-BFM-gamma}
\end{center}
\end{figure*}

Two models are used for simulating linear polymers in the bulk under
good solvent conditions. One is the standard SAW on the simple cubic
lattice, effective monomers being described by occupied lattice sites,
connected by bonds of fixed length $\mid \vec{b} \mid =\ell_b=1$.
Each site can be visited only once, and thus the excluded volume
interaction is realized. The other is the standard bond fluctuation 
model (BFM).
On the simple cubic lattice
each effective monomer of a SAW chain blocks all
eight corners of an elementary cube of the lattice
from further occupation. Two successive monomers along a chain are 
connected by a bond vector $\vec{b}$ which is taken from the set
\{$(\pm2,0,0)$,$(\pm 2,\pm 1,0)$,
$(\pm 2, \pm 1, \pm 1)$, $(\pm 2, \pm 2, \pm 1)$, $(\pm 3, 0,0)$,
$(\pm 3,\pm 1,0)$\} including also all permutations. The bond length
$\ell_b$ is therefore in a range between $2$ and $\sqrt{10}$.
There are in total
$108$ bond vectors and $87$ different bond angles between two sequential bonds
along a chain serving as candidates for building the conformational
structure of polymers.
The partition sum of a SAW of $N$ steps is 
\begin{equation}
        Z_N=\sum_{\rm config.} 1 
\end{equation}
which is simply the total number of possible configurations consisting 
of $(N+1)$ monomers.

  In the literature there are still no estimates of the fugacity 
$\mu_\infty(=1/q_{\rm eff})$ and the entropic exponent $\gamma$ for
SAWs on the BFM. According to the scaling law of the partition sum $Z_N$,
Eq.~(\ref{eq-ZN}), the  effective entropic exponent
$\gamma_{\rm eff}^{(1)}$ obtained from triple ratios~\cite{Grassberger1997b}
\begin{equation}
     \gamma_{\rm eff}^{(1)}(N) = 1+\frac{7\ln Z_N - 6 \ln Z_{N/3} - \ln Z_{5N}}
{\ln (3^6/5)}
\label{eq-gamma1}
\end{equation}
is shown in Fig.~\ref{fig-BFM-gamma}a. 
It gives $\gamma=\lim_{N \rightarrow \infty} \gamma_{\rm eff}^{(1)}(N)=1.1578(6)$. 
The fugacity $\mu_\infty$ is therefore determined by adjusting the value $a$
such that the curve of ${\ln Z_N + N \ln a-(\gamma-1) \ln N}$ with 
$\gamma=1.1578$
becomes horizontal for very large $N$ (see Fig.~\ref{fig-BFM-gamma}b).
We obtain the fugacity ${\mu_\infty=0.0117241395(75)}$
and the corresponding 
effective coordination number ${q_{\rm eff}=85.294106(55)}$ listed in Table~\ref{table1}.
In Fig.~\ref{fig-BFM-gamma}a we also show the asymptotic behavior
of the effective entropic exponent $\gamma_{\rm eff}^{(2)}$ defined by
\begin{equation}
  \gamma_{\rm eff}^{(2)}(N)= 1+ 
\frac{\ln \left[\mu^{3N/2}Z(2N)/Z(N/2)\right]}{\ln 4}
\label{eq-gamma2}
\end{equation}
with our estimate of $\mu_\infty$ for comparison.

If the excluded volume effect is ignored completely, a polymer chain behaves as
an ideal chain. It is well described by a random walk (RW), a walk that
can cross itself or may trace back the same path, or by a non-reversal
random walk (NRRW) where immediate back tracing is not allowed.
The partition sums of RW and NRRW are given by 
\begin{equation}
    Z_N = q^{N} \quad {\rm (RW)}\,, \quad Z_N = q(q-1)^{N-1} 
\quad {\rm (NRRW)} 
\end{equation}
where $q$ is the coordination number. $q=6$ for the standard
RW on the simple cubic lattice, and $q=108$ for the BFM~\cite{Kremer1988}.
The Flory exponent is $\nu=1/2$ for an ideal chain and
its mean square gyration radius 
$\langle R_g^2 \rangle= \langle R_e^2 \rangle /6$.

For the simulations of single RW, NRRW, and SAW chains we use the pruned-enriched
Rosenbluth method (PERM)~\cite{Grassberger1997}. It is a biased
chain growth algorithm with resampling and population control.
In this algorithm a polymer chain is built like a random walk by
adding one monomer at each step with a bias depending
on the problem at hand, and each configuration carries its own weight. 
The population control at each step is made such that
the ``bad" configurations are pruned with a certain probability,
and the ``good" configurations are enriched by properly reweighting,
until a chain has either grown to the maximum length of steps, $N$, or has
been killed due to attrition.
A detailed description of the algorithm PERM and its applications
is given in a review paper~\cite{HsuR2011}.
The algorithm has the advantage that the partition
sum can be estimated very precisely and directly.
It is also very efficient
for simulating linear polymer chains up to very long chain lengths
in dilute solution at and above the $\Theta$-point.
Therefore, we apply
the algorithm on the two lattice models, SCLM and BFM, 
in order to check for major differences between 
these two microscopic models.
{The longest chain length is $N=50000$ in our simulations here.}

\begin{figure*}[t]
\begin{center}
(a)\includegraphics[scale=0.29,angle=270]{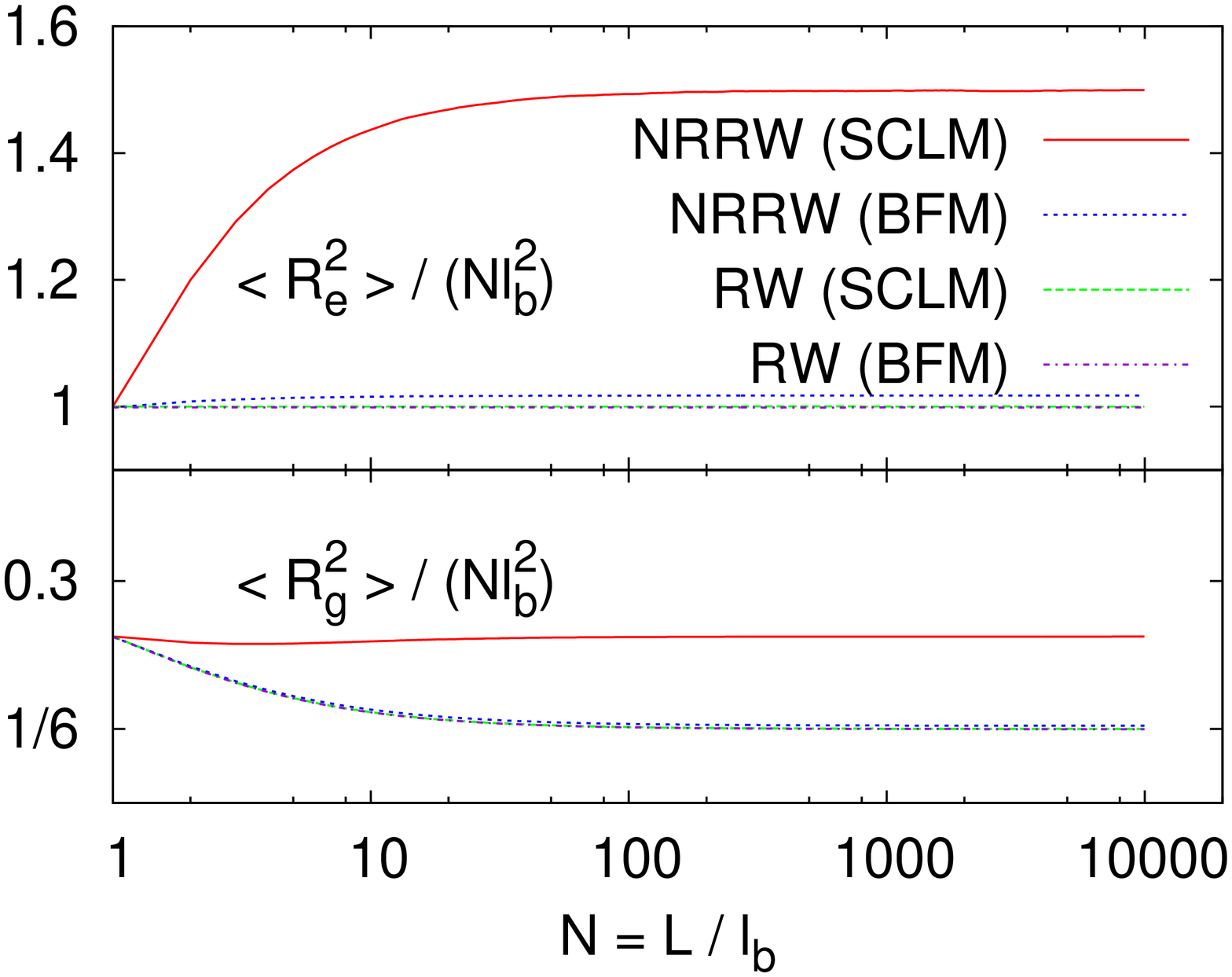} \hspace{0.4cm}
(b)\includegraphics[scale=0.29,angle=270]{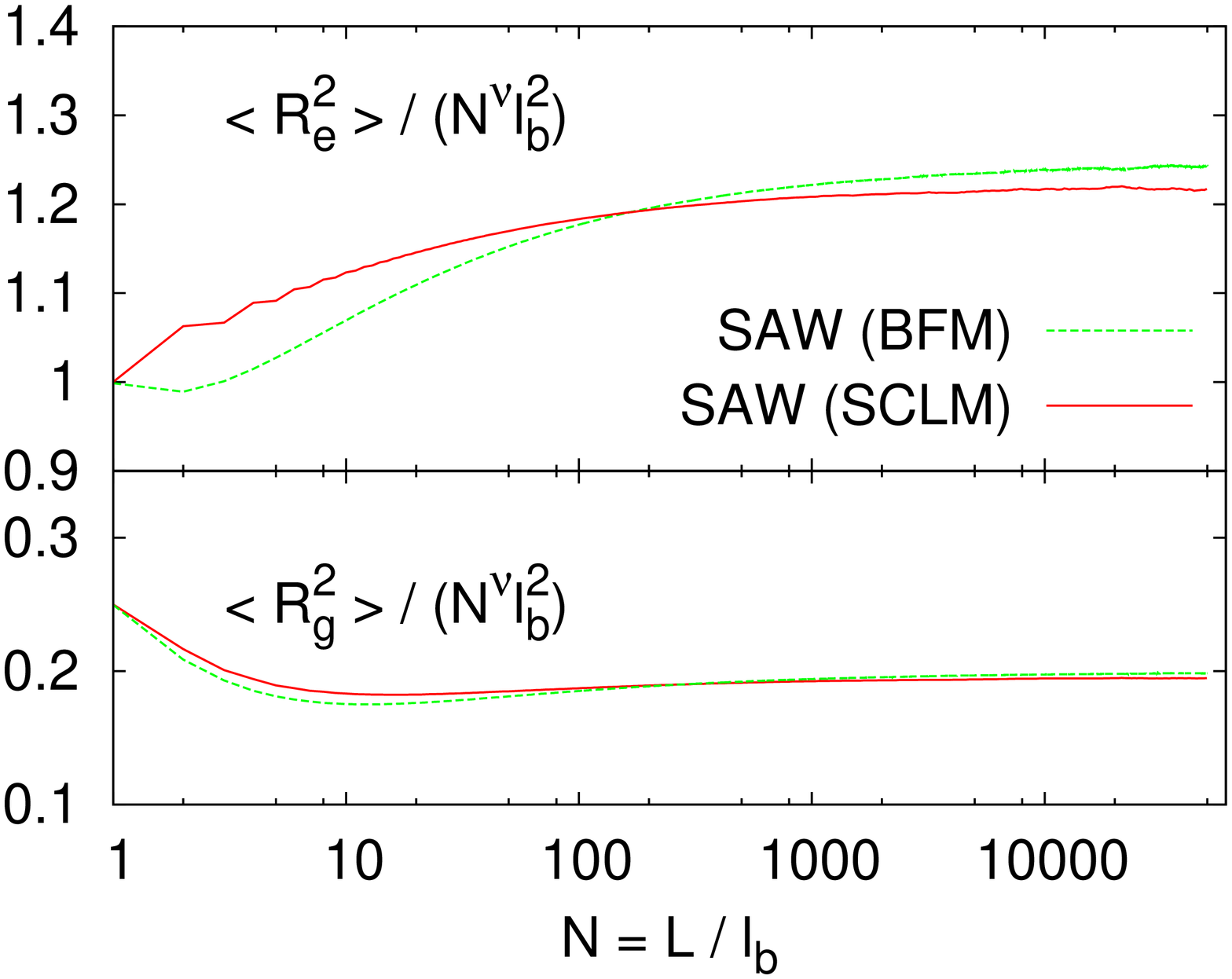} \hspace{0.4cm}
\caption{Mean square end-to-end distance $\langle R^2_e \rangle$ and
gyration radius $\langle R^2_g \rangle$ scaled by $(N^\nu \ell_b^2)$
with the Flory exponent $\nu=1/2$ for RWs and NRRWs (a), and
$\nu=0.5876$ for SAWs~\cite{Clisby2010}, plotted against $N$.
Here the bond length $\ell_b = \langle \vec{b}^2 \rangle^{1/2}$:
$\ell_b=1$ (SCLM) and $\ell_b=2.72$ (BFM).}
\label{fig-R2}
\end{center}
\end{figure*}

\begin{figure*}[t]
\begin{center}
(a)\includegraphics[scale=0.29,angle=270]{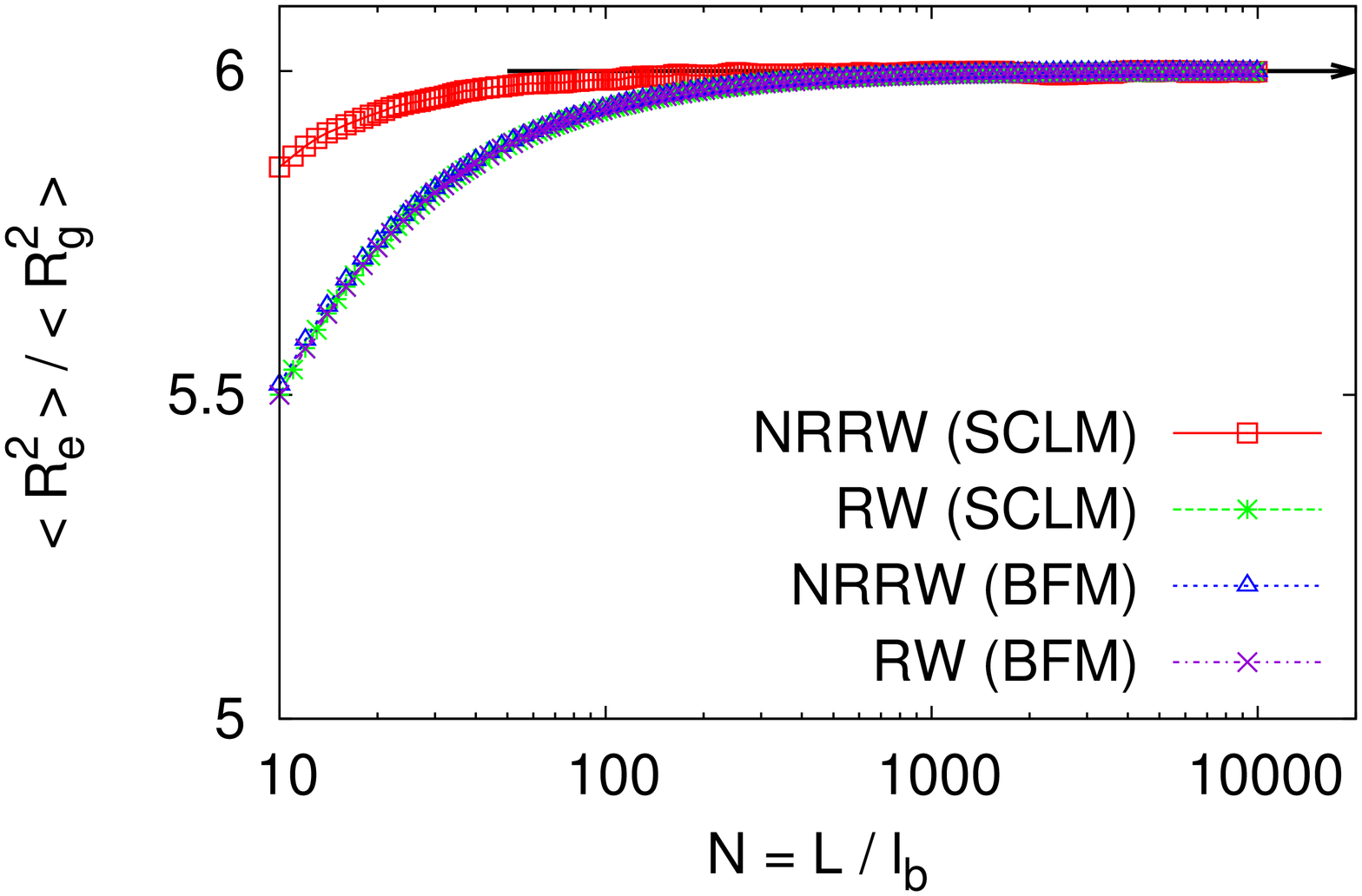}
(b)\includegraphics[scale=0.29,angle=270]{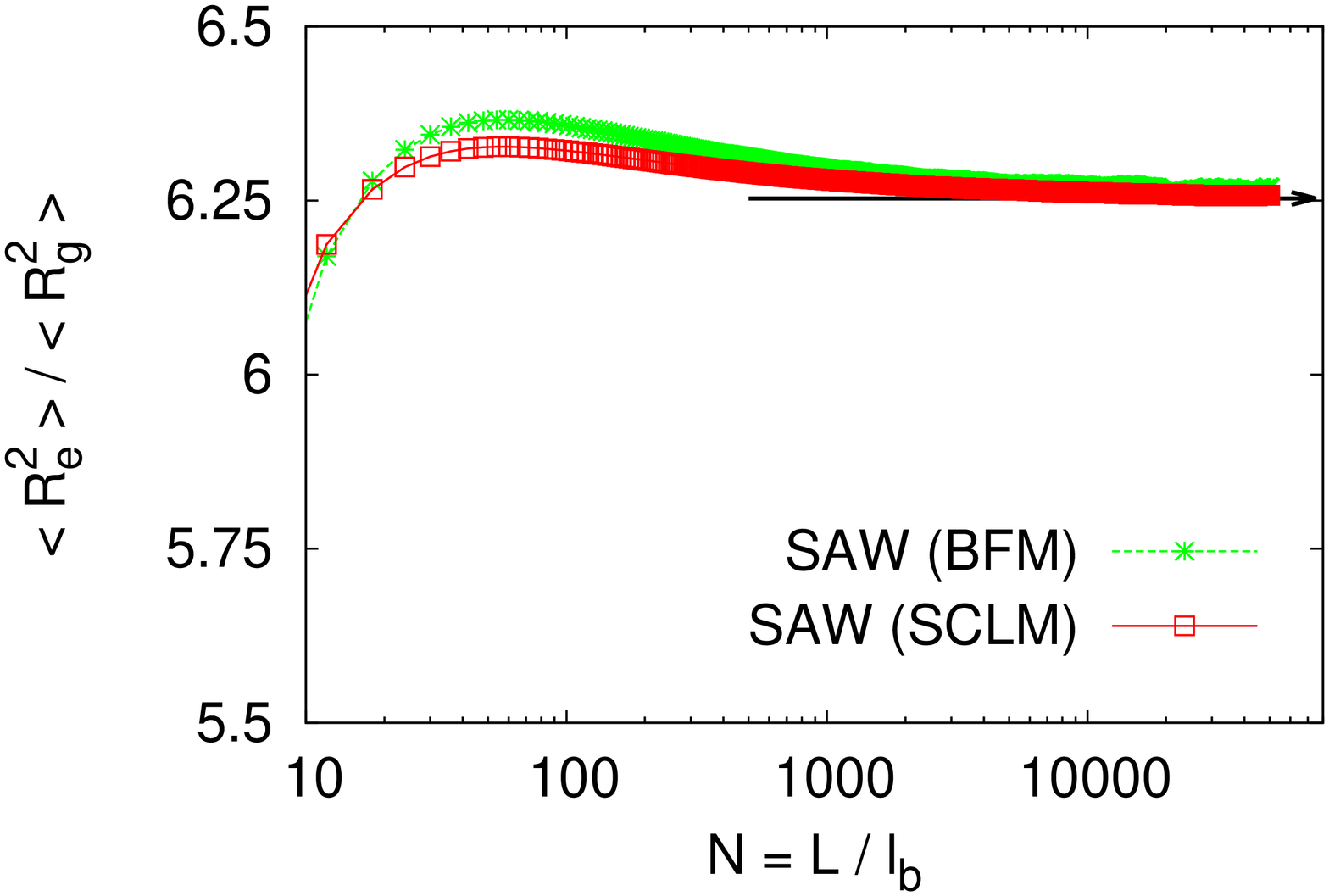}
\caption{Ratio between the mean square end-to-end distance and
the mean square gyration radius, $\langle R^2_e \rangle/\langle R^2_g \rangle$,
 plotted against the chain segments $N$, for RWs and NRRWs (a),
and for SAWs (b). As $N \gg 0$, $\langle R_e^2 \rangle/\langle R_g^2 \rangle \sim 6.00$ in (a)
and $\langle R_e^2 \rangle/\langle R_g^2 \rangle \sim 6.25$ in (b).}
\label{fig-ratio}
\end{center}
\end{figure*}

\begin{table*}[t]
\caption{The estimates of fugacity $\mu_\infty$, the effective
coordination number $q_{\rm eff}$ in Eq.~(\ref{eq-ZN}),
the amplitudes $A_e$ and $A_g$ in Eqs.~(\ref{eq-Re}) and (\ref{eq-Rg})
determined from the simulation data of $\langle R_e^2 \rangle$ and
$\langle R_g^2 \rangle$ for RWs, NRRWs, and SAWs based on the two lattice
models, SCLM and BFM.}
\label{table1}
\begin{center}
\begin{tabular}{|c|ccc|ccc|}
\hline
& \multicolumn{3}{|c|}{SCLM} & \multicolumn{3}{c|}{BFM} \\
\hline
model & RW & NRRW & SAW & RW & NRRW & SAW \\
\hline
$\mu_\infty$ & 1/6  & 1/5  & 0.21349098(5)~\cite{Grassberger2005} & 1/108 & 1/107 & 0.01172414395(75) \\
$q_{\rm eff}$ & 6 & 5  & 4.6840386(11) & 108  & 107  & 85.294106(55) \\
$A_e$ & 1.0000(2) & 1.4988(4) & 1.220(3) & 0.9986(2) & 1.0714(2) & 1.247(5) \\
$A_g$ & 0.16666(0) & 0.24985(7) & 0.1952(4) & 0.16645(4) & 0.16959(3) & 0.1993(6) \\
\hline
\end{tabular}
\end{center}
\end{table*}

\begin{figure*}[t]
\begin{center}
(a)\includegraphics[scale=0.29,angle=270]{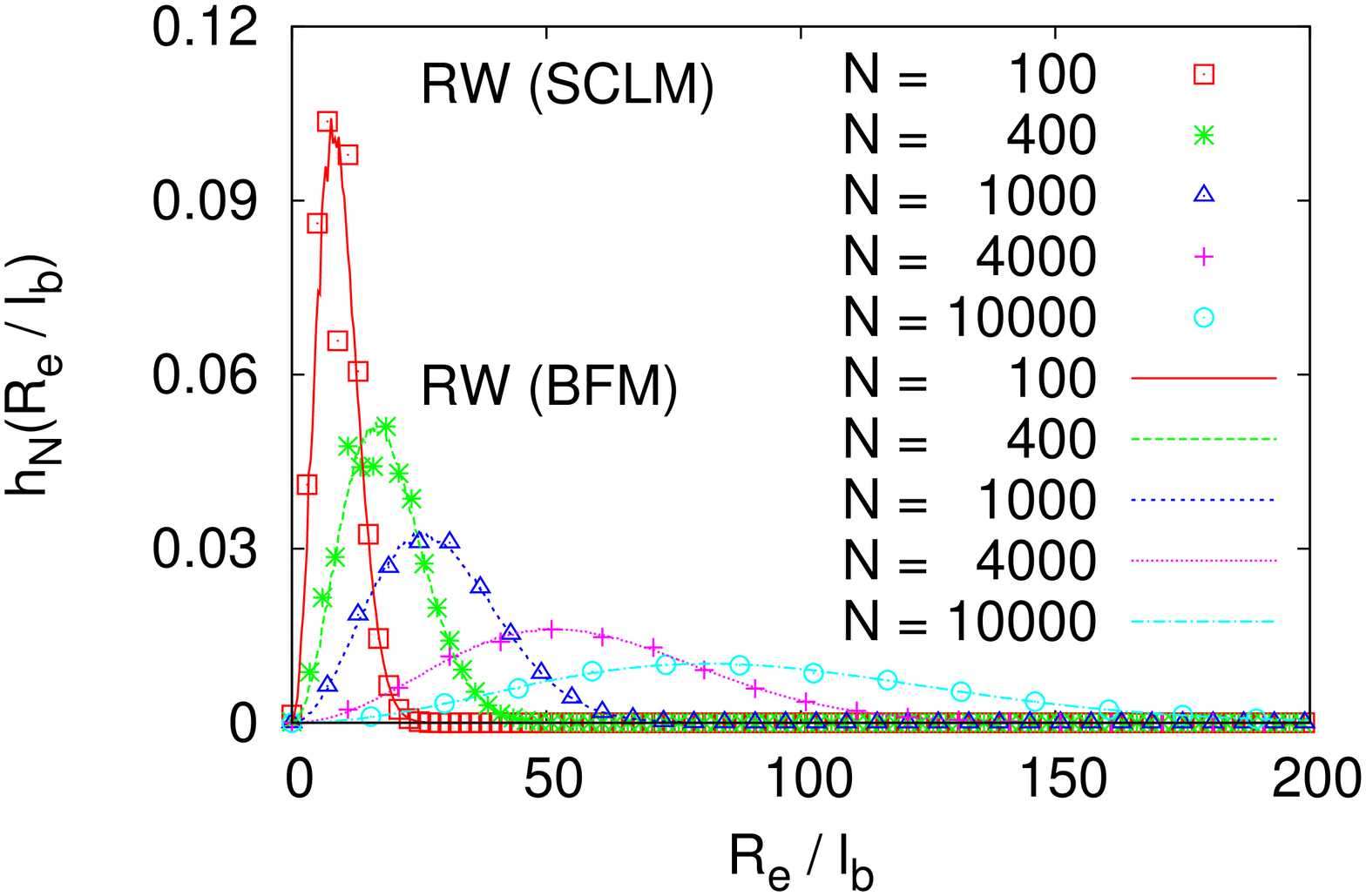} \hspace{0.4cm}
(b)\includegraphics[scale=0.29,angle=270]{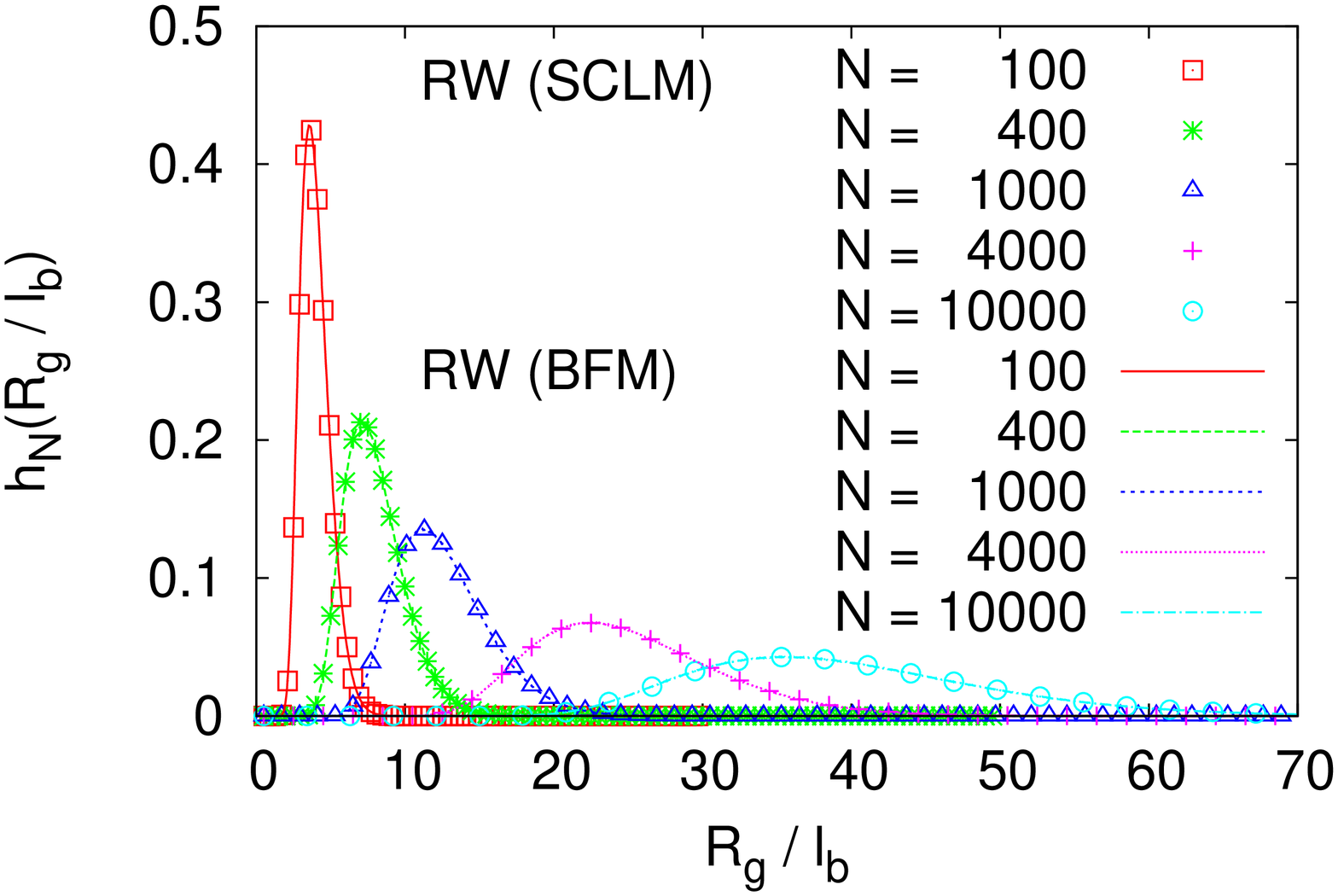}
(c)\includegraphics[scale=0.29,angle=270]{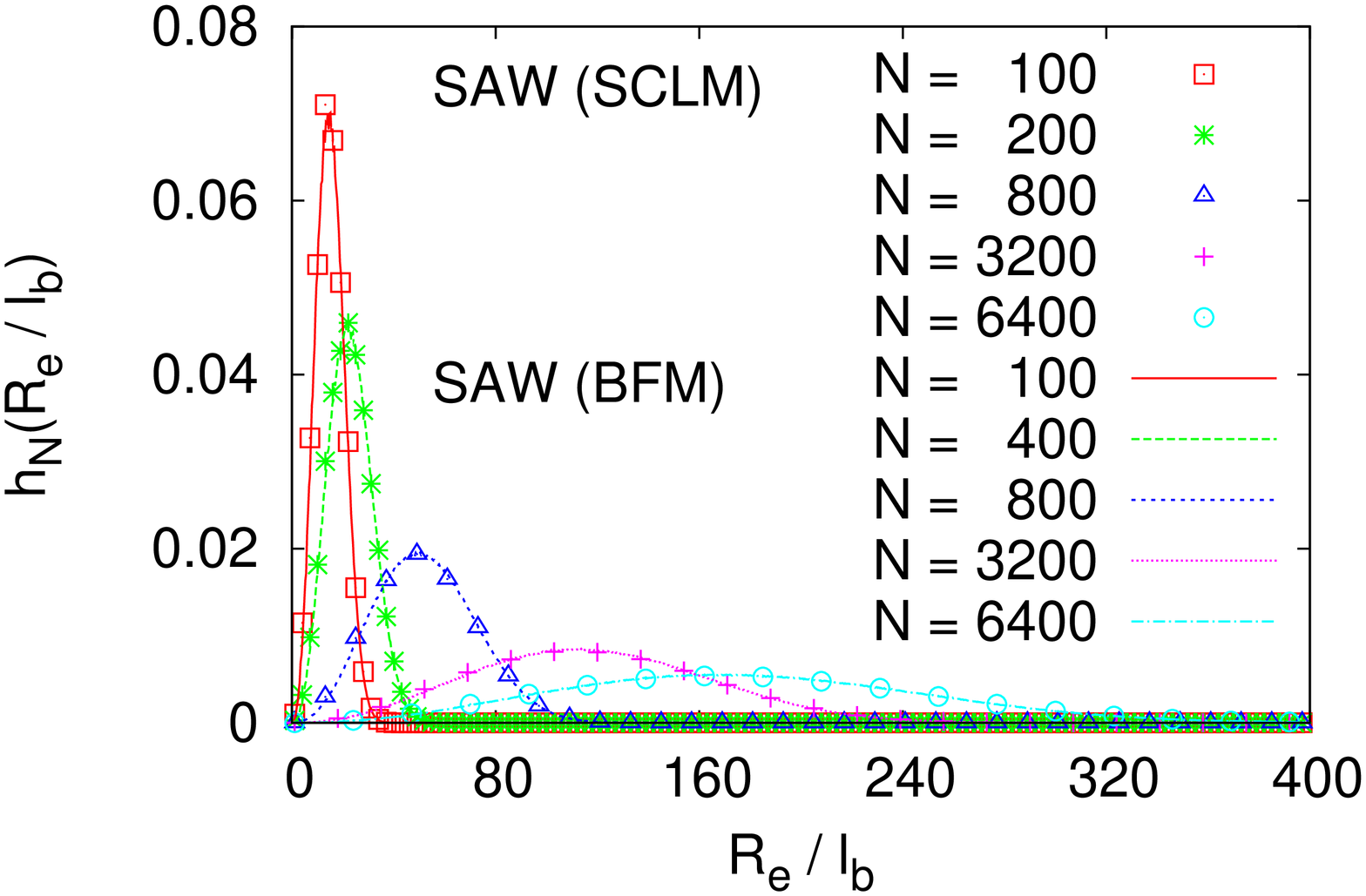} \hspace{0.4cm}
(d)\includegraphics[scale=0.29,angle=270]{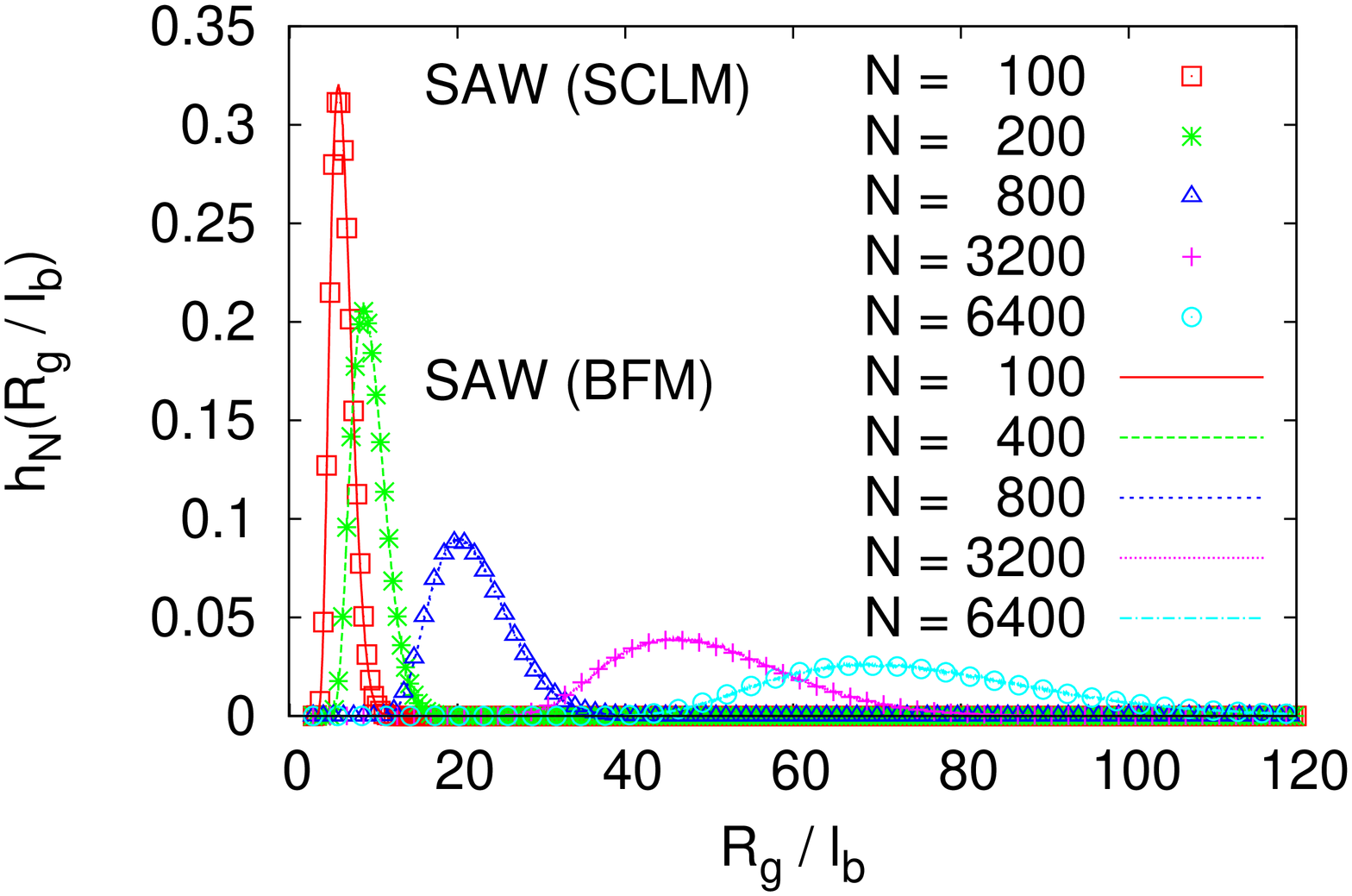}
\caption{(a),(c) Normalized probability distributions of end-to-end distance,
$h_N(R_e/\ell_b)(={\cal P}(R_e/\ell_b))$, plotted versus $R_e/\ell_b$. (b),(d)
similar as (a),(c), but
for gyration radius $R_g$. Data are for RWs (a),(b) and SAWs (c),(d).
Several values of chain lengths $N$ are chosen, as indicated.}
\label{fig-pReg}
\end{center}
\end{figure*}

\section{Conformations of single linear polymer chains: RWs, NRRWs, and SAWs}
\label{single}
We employ the pruned-enriched Rosenbluth method (PERM) for
the simulations of long single linear polymer chains 
of chain lengths (segments) up to $N \sim {\cal O}(10^4)$, modeled by
RWs, NRRWs, and SAWs depending on the interactions between monomers.
Figure~\ref{fig-R2}a with $\nu=1/2$ and Fig.\ref{fig-R2}b with $\nu=0.5876$
show that the scaling
laws, Eqs.~(\ref{eq-Re}) and (\ref{eq-Rg}), are verified as one should
expect. 
The mean square end-to-end distance simply is
\begin{equation}
       \langle R_e^2 \rangle =
\langle (\vec{r}_N-\vec{r}_0)^2 \rangle =\left \langle 
\left(\sum_{j=1}^N \vec{b}_j \right)^2 \right \rangle  \,.
\end{equation}
The mean square gyration radius is given by
\begin{eqnarray}
  \langle R_g^2 \rangle &=& \frac{1}{N+1}\left \langle 
  \sum_{j=0}^{N} (\vec{r}_j -\vec{r}_{CM} )^2 \right \rangle  \nonumber \\
  &=& \frac{1}{(N+1)^2} \left \langle \sum_{j=0}^N \sum_{k=0}^N
(\vec{r}_j -\vec{r}_k)^2 \right \rangle \, ,
\end{eqnarray}
where $\vec{r}_{CM}=\sum_{j=0}^N \vec{r}_j/(N+1)$ is the center of mass position
of the polymer.
The amplitudes $A_e$ and $A_g$ for RWs, NRRWs, SAWs based on
the two lattice models, SCLM and BFM, are listed in Table~\ref{table1}. 
Results of $\langle R_e^2 \rangle/(N\ell_b^2)$ and
$\langle R_g^2 \rangle/(N\ell_b^2)$ for RWs from both models follow
the same curves although the bond vectors in the BFM are not all
along the lattice directions and {do not} have the same bond length.
Here $\ell_b$ is the root-mean square bond
length, $\ell_b=1$
for the SCLM and $\ell_b=2.72$ for the BFM.
In Fig.~\ref{fig-R2}a, values of $\langle R_e^2 \rangle/(N\ell_b^2)$ and
$\langle R_g^2 \rangle/(N\ell_b^2)$ for NRRWs, obtained from SCLM 
for all lengths $N$ are significant larger than that obtained from the BFM,
since at each step the walker can only go straight or make a $90^o$ L-turn
in the SCLM.
In Fig.~\ref{fig-R2}b, the two
curves showing the results of $\langle R_e^2 \rangle/(N^\nu \ell_b^2)$
[$\langle R_g^2 \rangle/(N^\nu \ell_b^2)$] with $\nu=0.5876$
as functions of $N$ obtained from the two models 
intersect at $N \approx 180$,
and finally the amplitude for BFM is larger in the asymptotic regime.  
The slight deviation from the plateau value
is due to the finite size effects. The correction exponent $\Delta$
[Eqs.~(\ref{eq-Re}) and (\ref{eq-Rg})]
for these two models is determined by plotting
$\langle R_e^2 \rangle/N^{2\nu}$ and $\langle R_g^2 \rangle/N^{2\nu}$
versus $x \equiv N^{-\delta}$ (not shown). One should expect straight
lines near $x=0$ if and only if $\delta=\Delta$.
We obtain $\Delta=0.48(5)$ for both models, which is in agreement
with the previous simulation in Ref.~\cite{Li1995,Grassberger1997b,Clisby2010}
within error bars. 
The ratio between the mean square
end-to-end distance and the mean square gyration radius, is indeed
$\langle R_e^2 \rangle /\langle R_g^2 \rangle \approx  6.000(3)$
for RWs and NRRWs (Fig.~\ref{fig-ratio}a). For SAWs our results
give $\langle R_e^2 \rangle /\langle R_g^2 \rangle = 6.25(2)$
(Fig.~\ref{fig-ratio}b).
For SAWs on the simple cubic lattice the most
accurate estimates of $A_e=1.22035(25)$, $A_g=0.19514(4)$,
and $A_e/A_g = 6.2537(26)$ are given
in Ref.~\cite{Clisby2010}. Our results are also in perfect agreement
with them.
However, much longer chain lengths will be needed for a more 
precise estimate of 
the plateau value of the ratio in the asymptotic scaling regime.
Note that for $10 < N < 100$ the behavior is clearly model-dependent. 

We include here the RW and NRRW versions of both models not just for 
the sake of an exercise: often the mapping from an atomistic to a 
coarse-grained model is to be done under melt conditions, where excluded 
volume interactions are screened.

The shapes of polymer chains can also be described by the probability distributions
of end-to-end distance and gyration radius, $P(R_e/\ell_b)$ and $P(R_g/\ell_b)$,
respectively.
Numerically, they are obtained by accumulating the histogram $H_N(x)$ of $x$
over all configurations of length $N$, given by
\begin{equation}
    H_N(x) = \sum_{\rm configs.} W_N(x')\delta_{x,x'} \,,
\end{equation}
here each configuration carries its own weight $W_N(x')$. 
The normalized histogram is therefore,
\begin{equation}
 h_N(x)=H_N(x)/\sum_{x'} H_N(x') \, .
\end{equation}
Results of $h_N(R_e/\ell_b)$ and $h_N(R_g/\ell_b)$
for RWs and SAWs obtained from the 
two models for various values of chain lengths $N$
are shown in Fig.~\ref{fig-pReg}.
We see that both models give for $N>100$ the same distributions
of $R_e/\ell_b$ and $R_g/\ell_b$ although the mean values of
$R_e/\ell_b$ and $R_g/\ell_b$ are slightly different between
these two lattice models (Fig.~\ref{fig-R2}b) for SAWs.
Note that an angular average over all directions has been
included in the accumulating process of the histogram due to
spherical symmetry. Thus,
the normalized histograms of $R_e/\ell_b$,
\begin{equation}
 h_N\left(R_e/\ell_b \right) =
 {\cal P}_N({R_e}/{\ell_b}) =
4\pi C_{e,N} \left({R_e}/{\ell_b}\right)^2 P_N \left({\vec{R}_e}/{\ell_b} \right) 
\, ,
\label{eq-pRe-h}
\end{equation}
with
\begin{equation}
\int_0^\infty {\cal P}_N({R_e}/{\ell_b}) 
d\left(R_e/\ell_b\right) =1 \, ,
\label{eq-pRe-c}
\end{equation}
and the normalized histograms of $R_g/\ell_b$,
\begin{equation}
 h_N\left(R_g/\ell_b \right) =
 {\cal P}_N({R_g}/{\ell_b}) =
4\pi C_{g,N}\left({R_g}/{\ell_b}\right)^2 P_N \left({R_g}/{\ell_b} \right) 
\, ,
\label{eq-pRg-h}
\end{equation}
with
\begin{equation}
\int_0^\infty {\cal P}_N({R_g}/{\ell_b}) 
d\left(R_g/\ell_b\right) =1 \, ,
\label{eq-pRg-c}
\end{equation}
where $C_{e,N}$ and $C_{g,N}$ are the normalization factors.

\begin{figure}[t]
\begin{center}
\includegraphics[scale=0.29,angle=270]{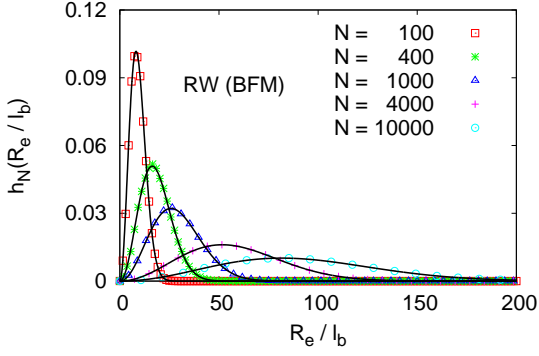}
\caption{Same as in Fig.~\ref{fig-pReg}a, but data are only for BFM.
The predicted distribution $4 \pi (R_e/\ell_b)^2P(R_e/\ell_b)$ 
for various values of $N$ are shown by solid curves.
Here the distribution function $P(R_e/\ell_b)$ is given by 
Eq.~(\ref{eq-pRe-rw}).}
\label{fig-pRe-rw}
\end{center}
\end{figure}

The probability distribution of end-to-end distance for ideal chains
is simply a Gaussian distribution,
\begin{equation}
  P(\vec{R_e}/\ell_b)=\frac{1}{(2\pi N/3)^{3/2}} 
\exp(-\frac{3(R_e/\ell_b)^2}{2N}) \,.
\label{eq-pRe-rw}
\end{equation}
Our numerical data for RWs obtained from BFM and SCLM shown in 
Fig.~\ref{fig-pReg} are in perfect agreement with the Gaussian 
distribution (see Fig.~\ref{fig-pRe-rw}). From Eqs.~(\ref{eq-pRe-rw}),
(\ref{eq-pRe-h}) and (\ref{eq-pRe-c}) we obtain the normalized
factor $C_{e,N}=1$. 

\begin{figure}[t]
\begin{center}
\includegraphics[scale=0.29,angle=270]{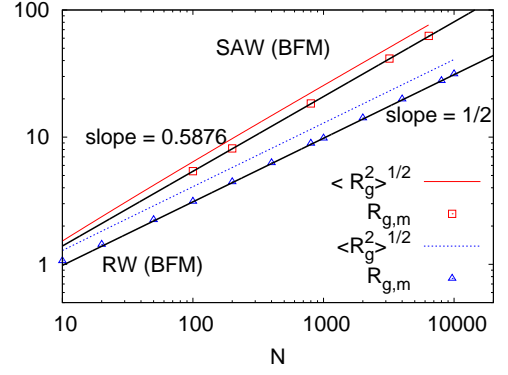}
\caption{Root mean square radius of gyration, $\langle R_g^2 \rangle^{1/2}$,
and the gyration radius $R_{g,m}$ where $P(R_g)$ has its maximum value,
plotted against chain length $N$ for RWs and SAWs. Data are for BFM.}
\label{fig-Rgm}
\end{center}
\end{figure}

\begin{figure*}[t]
\begin{center}
(a)\includegraphics[scale=0.29,angle=270]{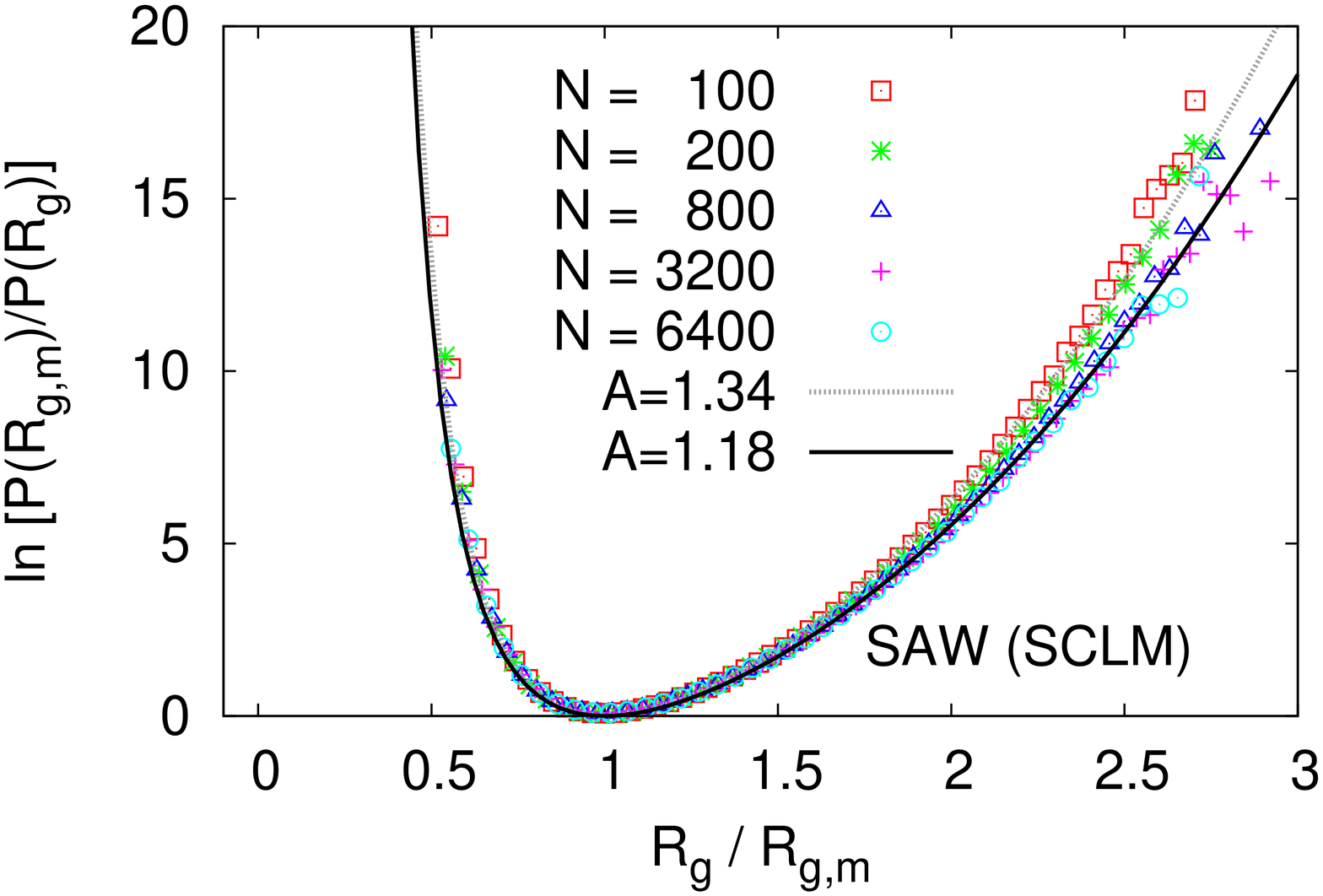} \hspace{0.4cm}
(b)\includegraphics[scale=0.29,angle=270]{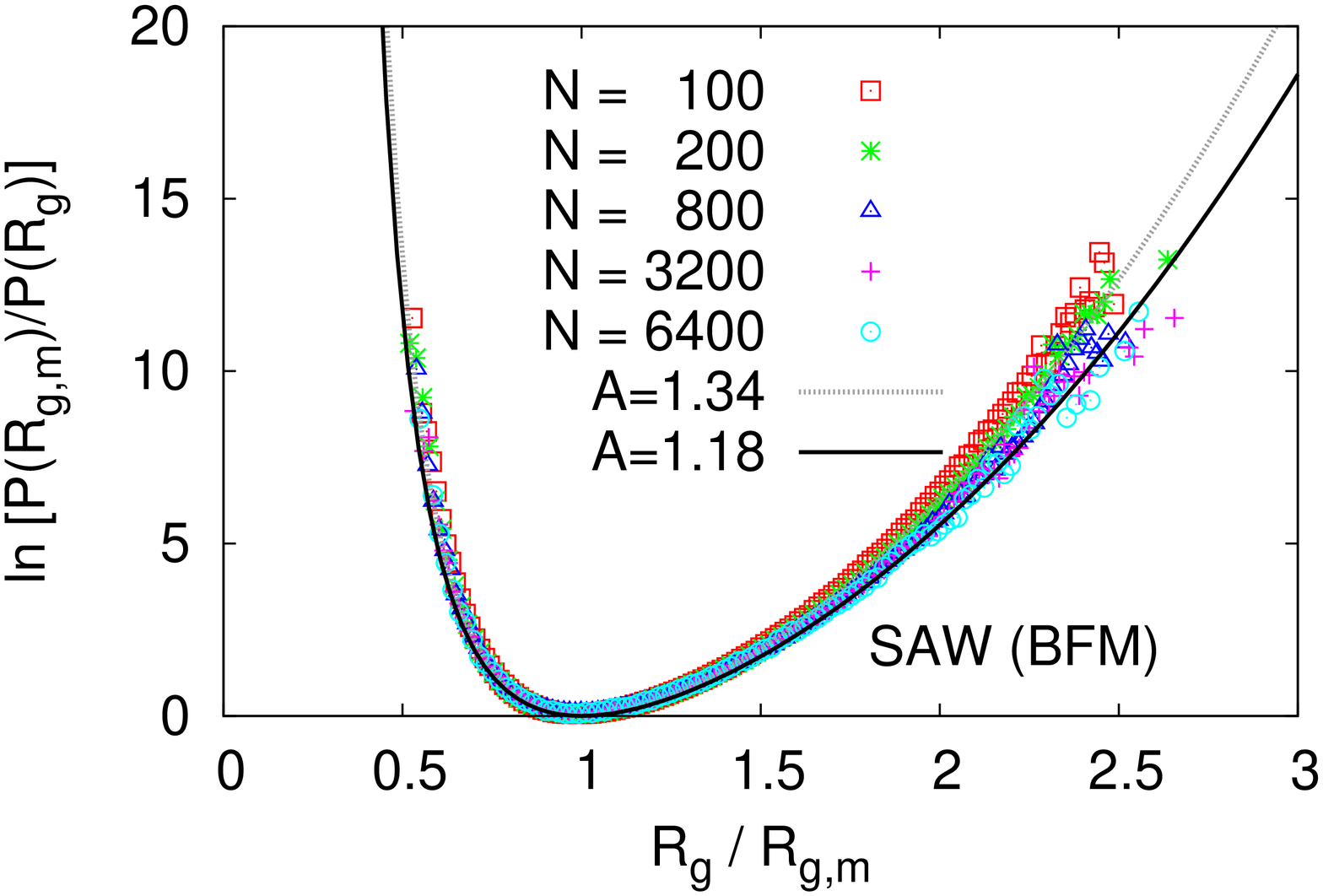}
\caption{Logarithm of the rescaled probability distribution of gyration radius,
$\ln(P(R_{g,m})/P(R_g))$ as a function of $R_g/R_{g,m}$ for SAWs
obtained from the models (a) SCLM and (b) BFM.
The best fit of Eq.~(\ref{eq-A}) with $A=1.18$ to our data is shown
by the solid curve.
The dashed curve with $A=1.34$ given in {Ref.~\cite{Bishop1991}} is also
shown for the comparison. Several values of chain lengths $N$ are chosen,
as indicated.}
\label{fig-lprg-saw}
\end{center}
\end{figure*}

\begin{figure*}[t]
\begin{center}
(a)\includegraphics[scale=0.29,angle=270]{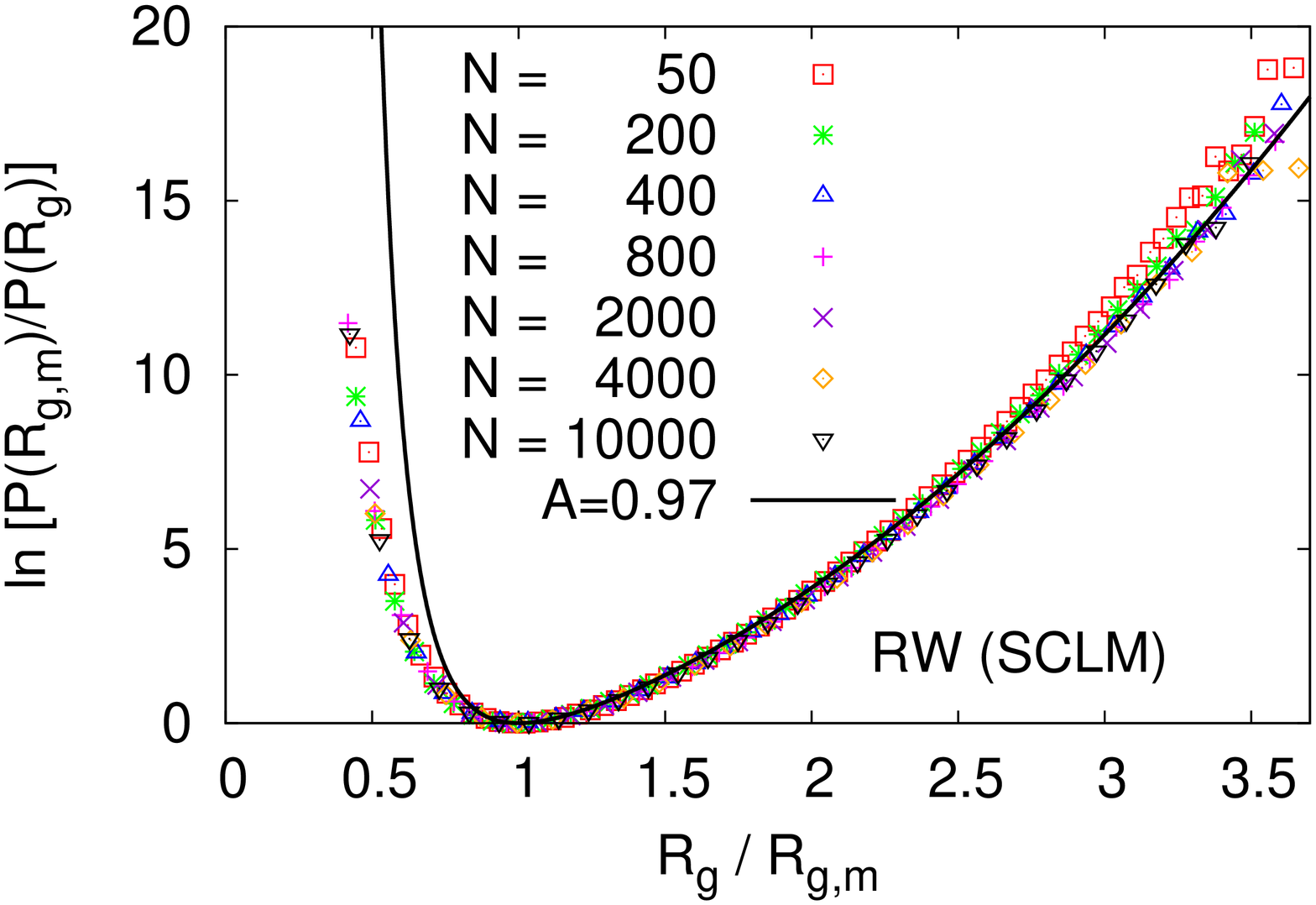} \hspace{0.4cm}
(b)\includegraphics[scale=0.29,angle=270]{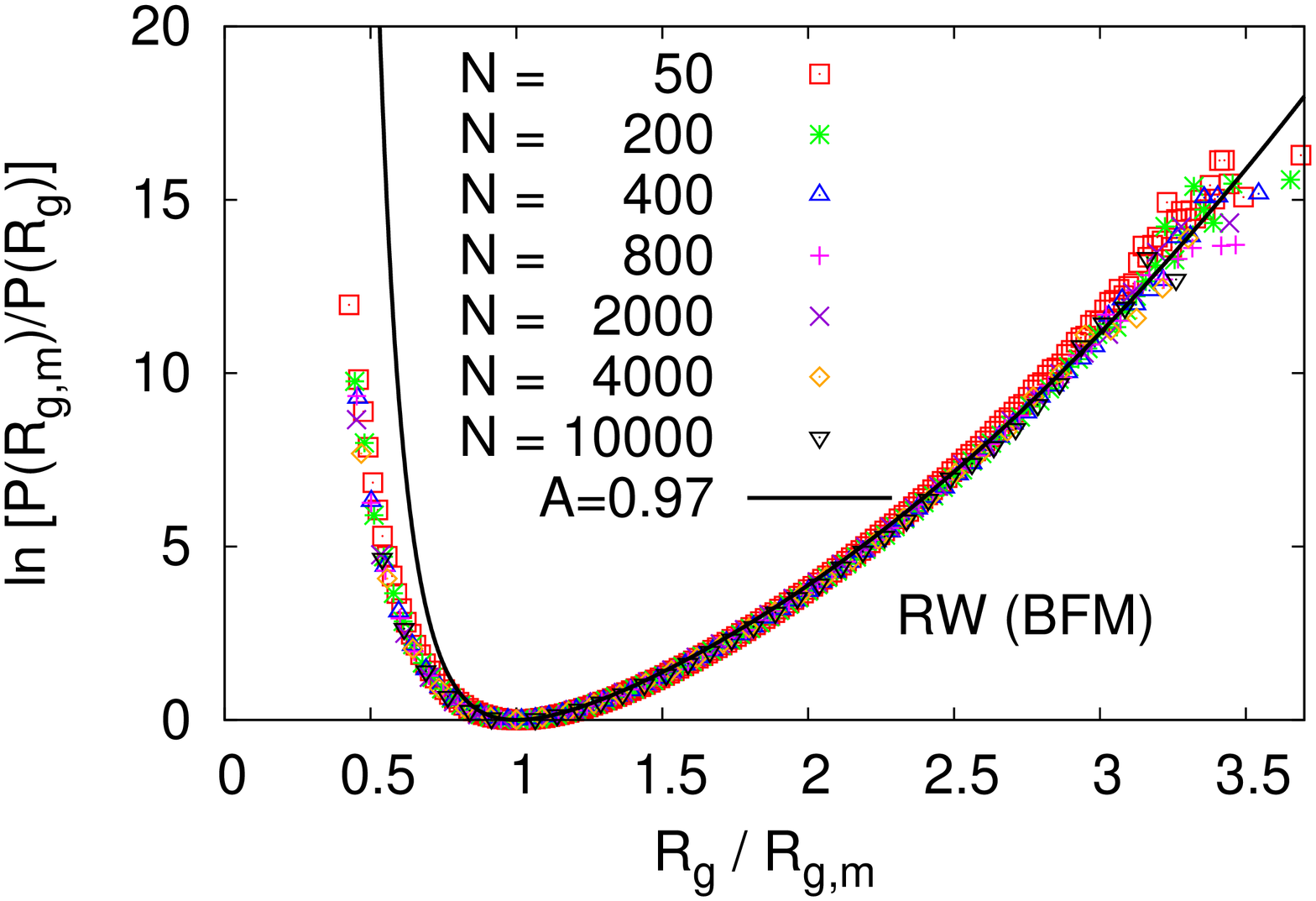}\\
\caption{Similar as in Fig.~\ref{fig-lprg-saw}, but for RWs.
The best fitting of our data gives $A=0.97$ for both models.}
\label{fig-lprg-rw}
\end{center}
\end{figure*}

\begin{figure*}[t]
\begin{center}
(a)\includegraphics[scale=0.29,angle=270]{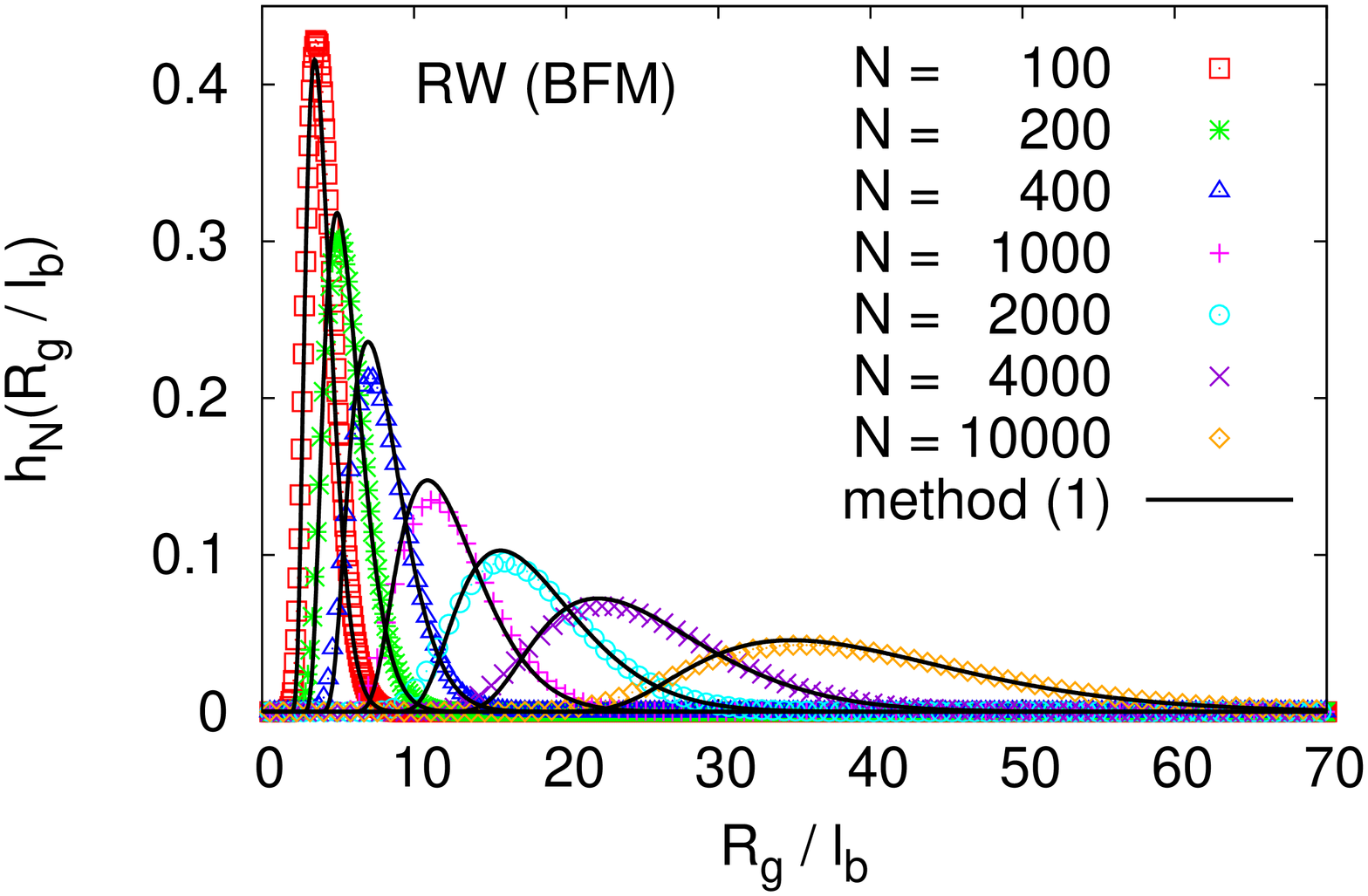} \hspace{0.4cm}
(b)\includegraphics[scale=0.29,angle=270]{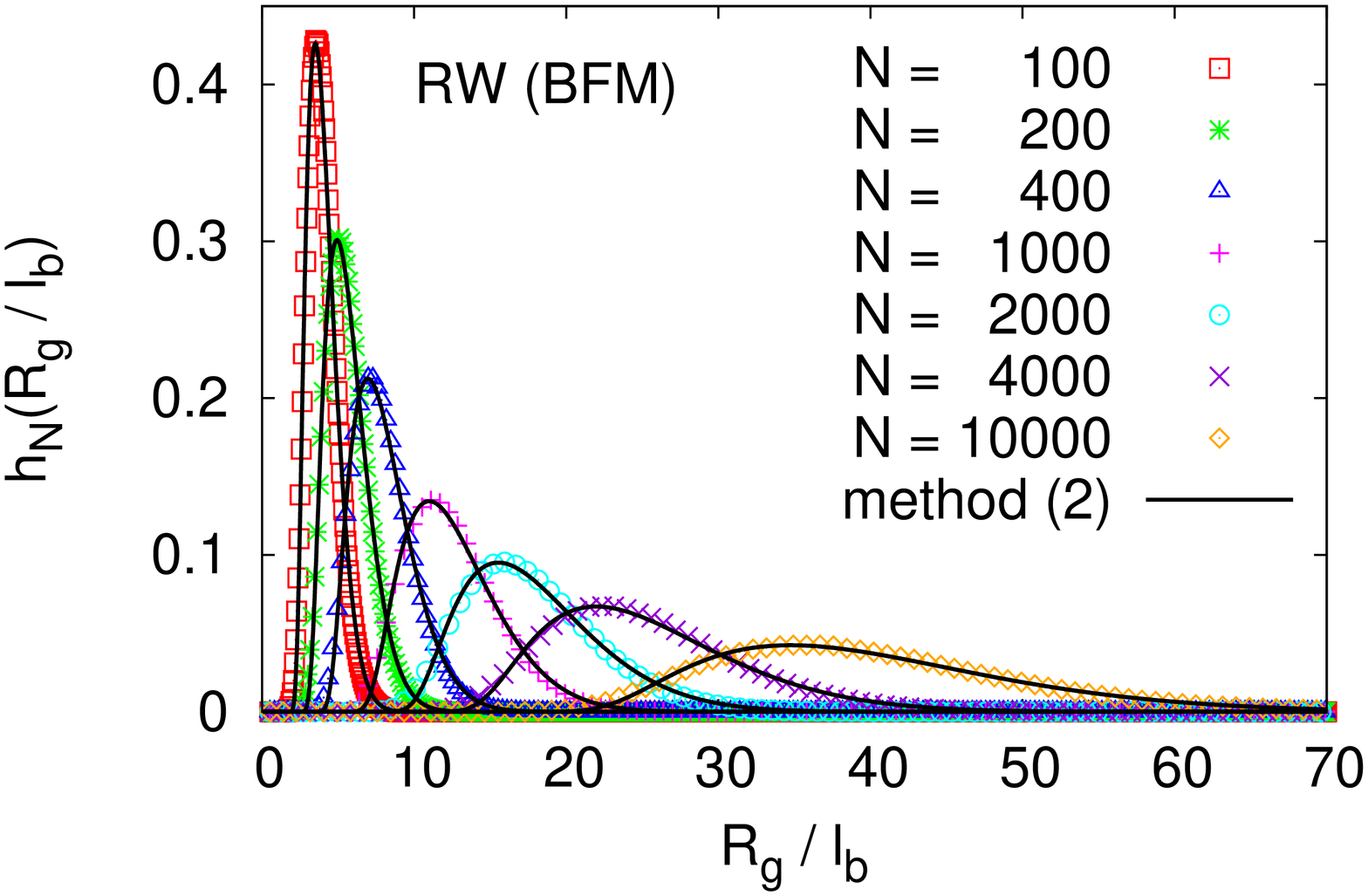}\\
\caption{Normalized probability distributions of $R_g/\ell_b$,
$h(R_g/\ell_b)={\cal P}_N(R_g/\ell_b)$, plotted versus $R_g/\ell_b$
for various values of $N$, and for BFM.
The fitting functions $4\pi C_{g,N}(R_g/\ell_b)^2 P(R_g/\ell_b)$
[Eqs.~(\ref{eq-pRg-h}), (\ref{eq-pRg-c}), and (\ref{eq-pRg})]
with parameters $a_1$, $a_2$, and $C_{g,N}$ determined by method (1)
and method (2)
are shown by curves for comparison in (a) and (b), respectively.}
\label{fig-prg-rw1}
\end{center}
\end{figure*}

The theoretical prediction of the gyration radius probability distribution
of polymer chains under good solvent conditions in $d$-dimensions 
suggested by Lhuillier~\cite{Lhuillier1988} is as follows:
\begin{equation}
  P(R_g/\ell_b) \sim \exp \left [- a_1 \left(\frac{\ell_bN^\nu}{R_g} \right)^{\alpha d}
- a_2\left(\frac{R_g}{\ell_bN^\nu} \right)^\delta \right]  
\label{eq-pRg}
\end{equation}
where $a_1$ and $a_2$ are (non-universal) constants, and the exponents
$\alpha$ and $\delta$ are linked to the space dimension $d$ and
the Flory exponent $\nu$ by
\begin{equation}
  \alpha=(\nu d -1)^{-1} \quad {\rm and} \quad \delta=(1-\nu)^{-1} \,.
\end{equation}
Here $(1+\alpha)$ is the des Cloizeaux exponent~\cite{Cloizeaux1975} 
for the osmotic pressure of a semidilute polymer solution,
and $\delta$ is the Fisher exponent~\cite{Fisher1966} characterizing
the end-to-end distance distribution.

This scaling form has been verified in the previous Monte Carlo simulation
studies of the standard self-avoiding walks on square ($d=2$) and cubic ($d=3$) lattices
up to ${\cal O}(10^2)$ steps using the slithering-snake and pivot
algorithms~\cite{Victor1990, Bishop1991}. The two
fitting parameters $a_1$ and $a_2$ are actually not independent
since at the position where the distribution $P(R_g)$ has its maximum value,
i.e. $P(R_g=R_{g,m})=\max\; P(R_g)$, the corresponding gyration radius
$R_{g,m} \propto R_g \propto N^\nu$ (see Fig.~\ref{fig-Rgm}).
Using Eq.~(\ref{eq-pRg}), the logarithm of the rescaled probability
is written as
\begin{eqnarray}
&& f \left( \frac{R_{g,m}}{R_g} \right) = \ln 
\frac{P(R_{g,m}/\ell_b)}{P(R_g/\ell_b)} \nonumber \\
&=& A\left[\frac{1}{\alpha} \left( \frac{R_{g,m}}{R_g} \right)^{\alpha d}
+\frac{d}{\delta} \left( \frac{R_g}{R_{g,m}}\right)^\delta+1-d \right]
\label{eq-A}
\end{eqnarray}
with
\begin{equation}
  a_1=\frac{A}{\alpha} \left(\frac{R_{g,m}}{\ell_bN^\nu} \right)^{\alpha d}
\quad {\rm and} \quad a_2=\frac{Ad}{\delta} 
\left(\frac{\ell_b N^\nu}{R_{g,m}} \right)^\delta  \,.
\end{equation}
From Eq.~(\ref{eq-pRg-c}), we obtain
\begin{equation}
     \ln \frac{P(R_{g,m}/\ell_b)}{P(R_g/\ell_b)} = 
\ln \frac{h_N(R_{g,m}/\ell_b)}{h_N(R_g/\ell_b)} + 2 \ln \frac{R_g}{R_{g,m}}
\, .
\end{equation}
Our estimate of $\ln \left(P(R_{g,m}/\ell_b)/P(R_g/\ell_b) \right)$ for SAWs based on the two
lattice models, SCLM and BFM, are shown in Fig.~\ref{fig-lprg-saw}.
As chain lengths $N>1000$, we see the nice data collapse, and
the logarithm of the scaled probability of $R_g$ is described by
Eq.~(\ref{eq-A}) with $A=1.18$ very well.
Due to the finite-size effect it is clearly seen that the previous estimate
$A=1.34$ is an overestimate~\cite{Bishop1991}.

\begin{figure}[t]
\begin{center}
\includegraphics[scale=0.29,angle=270]{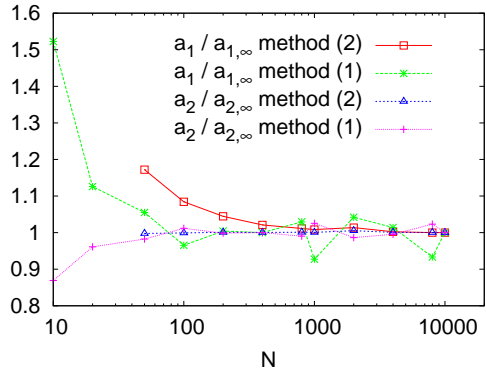} \hspace{0.4cm}
\caption{Parameters $a_1$ and $a_2$ in Eq.~(\ref{eq-pRg}) plotted 
versus $N$. Results are obtained from two different methods mentioned in the text.
Here $a_{1,\infty}$ and $a_{2,\infty}$ are taken from Table~\ref{table2} for $N=10000$.}
\label{fig-a12}
\end{center}
\end{figure}

For an ideal chain the distribution of $R_g$ is no longer a simple 
Gaussian distribution as shown in Eq.~(\ref{eq-pRe-rw}), and the
exact expression is quite complicated. Vettorel et al.~\cite{Vettorel2010}
found out the formula given by Lhuillier~\cite{Lhuillier1988} is 
a good approximation for describing the distribution $P(R_g)$ for an ideal
chain based on the BSM. Therefore, we also use 
the same formula for the investigation of the distribution $P(R_g)$
obtained from the two coarse-grained lattice models.
Two methods are discussed here. Method (1): We use the formula [Eq.~(\ref{eq-A})] 
which contains only one fitting parameter $A$ since $R_{g,m} \sim R_g\sim N$ 
for RWs as shown in Fig.~\ref{fig-Rgm}. From our simulations of RWs, we still 
see the nice data collapse for $N>200$ in the plot of the logarithm of the 
rescaled distribution of $R_g$ (Fig.~\ref{fig-lprg-rw}), but the distribution 
can only be described by Eq.~(\ref{eq-A}) well for $R_g>R_{g,m}$. Using the 
least square fit, it gives $A=0.97$. Method(2): We assume that the two 
parameters $a_1$ and $a_2$ in Eq.~(\ref{eq-pRg}) are independent. 
Using Eqs.~(\ref{eq-pRg-h}), (\ref{eq-pRg-c}), and (\ref{eq-pRg}),
values of $a_1$, $a_2$, and the normalization factor $C_{g,N}$
are determined by the best fit of the normalized histograms $h_N(R_g/\ell_b)$ 
obtained from our Monte Carlo simulations. Note that it is not
possible to determine $a_1$ and $a_2$ using the second method for $N<50$ since
the normalization condition, Eq.~(\ref{eq-pRg-c}), is not satisfied. 
In Fig.~\ref{fig-prg-rw1} we compare our  
results of $h_N(R_g/\ell_b)\propto {\cal P}(R_g/\ell_b) \propto (R_g/\ell_b)^2P(R_g/\ell_b)$
for BFM to the fitting function $4\pi C_{g,N} (R_g/\ell_b)^2 P(R_g/\ell_b)$ 
[Eqs.~(\ref{eq-pRg-h}), (\ref{eq-pRg-c}), and (\ref{eq-pRg})]
with parameters determined by these two different methods.
Values of $a_1$ and $a_2$ plotted
versus $N$ are shown in Fig.~\ref{fig-a12} and listed in Table~\ref{table2}.
Our results show that $a_1$ and $a_2$ are almost constants for large $N$,
which are comparable with the results obtained for the BSM~\cite{Vettorel2010}. 

\begin{table*}[t]
\caption{Parameters $a_1$ and $a_2$ of the probability distribution $P(R_g/\ell_b)$,
Eq.~(\ref{eq-pRg}), determined by two different methods (1) and (2) mentioned in
the text for various values of chain length $N$.} 
\label{table2}
\begin{center}
\begin{tabular}{|l|cccccccccccc|}
\hline
\multicolumn{1}{|c|}{$N$} & 10  & 20 & 50 & 100 & 200 & 400 & 800 & 1000 & 2000 & 4000 & 8000 & 10000\\
\hline
(1) $a_1 \times 10^4$ & 7.12 & 5.27 & 4.94 & 4.52 & 4.70 & 4.70 & 4.82 & 4.34 & 4.88 & 4.74 & 4.37 & 4.68\\
(1) $a_2 $ & 12.80 & 14.15 & 14.46 & 14.90 & 14.70 & 14.73 & 14.58 & 15.10 & 14.42 & 14.66 & 15.07 & 14.73\\
(2) $a_1 \times 10^4$ & \_ & \_ & 4.14 & 3.83 & 3.69 & 3.60 & 3.57 & 3.56 & 3.58 & 3.54 & 3.53 & 3.53 \\
(2) $a_2 $ & \_ & \_ & 13.25 & 13.27 & 13.29 & 13.29 & 13.30 & 13.29 & 13.35 & 13.29 & 13.29 & 13.28 \\
\hline
\end{tabular}
\end{center}
\end{table*}

\section{Semiflexible chains}
\label{semi}

We extend our simulations in this
section from flexible chains to semiflexible chains. 
Extensive Monte Carlo simulations of semiflexible polymer 
chains described by standard SAWs on the simple cubic lattice, with a 
bending potential 
$U_b=\varepsilon_b(1-\cos \theta)$, have been recently carried 
out~\cite{Hsu2010b,Hsu2010,Hsu2011}. 
Recall that atomistic models of real chains may exhibit considerable 
chain stiffness due to the combined action of torsional and bond angle 
potentials. When a mapping to a coarse-grained model is performed,
this stiffness is lumped into an effective bond angle potential of the 
coarse-grained model.
In this standard model the angle between two subsequent
bond vectors along the chain is either $0^o$ or $\pm 90^o$, and hence
in the statistical weight of a SAW configuration on the lattice every 
$90^o$ bend will contribute a Boltzmann factor $q_b=\exp(-\varepsilon_b/k_BT)$
($q_b=1$ for ordinary SAWs). $k_BT$ is of order unity throughout the whole paper.
The partition function of such a standard SAW
with $N$ bonds ($N+1$ effective monomers) and $N_{\rm bend}$ bends
is therefore,
\begin{equation}
      Z_N(q_b)=\sum_{\rm config.} C_N(N_{\rm bend}) q_b^{N_{\rm bend}} \,,
\label{eq-Zb-saw}
\end{equation}
where $C_N(N_{\rm bend})$ is the total number of all configurations of 
a polymer chain of length $N$ containing $N_{\rm bend}$ bends.

   We are also interested in understanding the microscopic difference between
the standard SAWs and the BFM as the stiffness of chains is taken into account.
Since there are $87$ bond angles possibly occurring in the chain conformations, 
the partition function cannot be simplified for the BFM, written as,
\begin{eqnarray}
&& Z_{N}^{\rm (BFM)} (\varepsilon_b) \nonumber \\
&=&\sum_{\rm config.} C_N(\{\theta\})
\exp\left[-\frac{\varepsilon_b}{k_BT}\sum_{i=1}^{N-1} 
(1-\cos \theta_{i,i+1}) \right] \,,
\label{eq-Zb-BFM}
\end{eqnarray}
where $\theta_{i,i+1}$ is the bond angle between the $i^{\rm th}$ bond vector
and the $(i+1)^{\rm th}$ bond vector along a chain, and
$C_N(\{\theta\})$ is the number of configurations having
the same set $\{\theta\}$ but fluctuating bond lengths.
In the absence of excluded volume effect,
the formulas of the partition function, Eq.~(\ref{eq-Zb-saw}) and 
Eq.~(\ref{eq-Zb-BFM}), 
remain the same while semiflexible chains are described by RWs and NRRWs. 

\subsection{Theoretical predictions}
There exist several theoretical models describing the behavior of 
semiflexible chains in the absence of excluded volume effects.
We first consider a discrete worm-like chain model~\cite{Winkler1994}
that a chain consisting of $N$ bonds
with fixed bond length $\ell_b$, but successive bonds are 
correlated with respect to their relative orientations,
\begin{equation}
   \langle \vec{b}_i \cdot  \vec{b}_{i+1} \rangle =\ell_b^2 
\langle \cos \theta \rangle \quad {\rm and} \quad
\langle \vec{b}_i^2 \rangle =\ell_b^2  \,,
\label{eq-cos-dWLC}
\end{equation}
where $\theta$ is the angle between the successive bond vectors. 
The mean square end-to-end distance is therefore,
\begin{equation}
   \langle R_e^2 \rangle =N\ell_b^2 \left[ 
\frac{1+\langle \cos \theta \rangle}{1- \langle \cos \theta \rangle}
-\frac{2 \langle\cos \theta \rangle( 1-(\langle \cos \theta \rangle)^N}
{N(1-\langle \cos \theta \rangle)^2} \right]  \, .
\label{eq-Re2-dWLC}
\end{equation}
This formula agrees with the prediction for a freely rotating chain (FRC).
In the limit $N\rightarrow \infty$ the bond-bond orientational
correlation function decays exponentially as a function of their
chemical distance $s$, 
\begin{equation}
 \langle  \vec{b}_i \cdot \vec{b}_{i+s} \rangle =\ell_b^2 
\langle \cos \theta(s) \rangle =\ell_b^2 \langle \cos \theta \rangle^s
=\ell_b^2 \exp(-s \ell_b / \ell_p) \,,
\label{eq-cos-dWLC1}
\end{equation}
where $\ell_p$ is the so-called persistence length which can be
extracted from the initial decay of $\langle \cos \theta (s) \rangle$.
Equivalently, one can calculate the persistence length from
\begin{equation}
    \ell_{p,\theta}/\ell_b = -1/\ln (\langle \cos \theta \rangle) 
\label{eq-lp-dWLC}
\end{equation}
here instead of $\ell_p$ we use $\ell_{p,\theta}$ to distinguish between
these two measurements.

For rather stiff ($L \gg \ell_p$) and long chains ($N \rightarrow \infty$)
we expect that the bond angles $\theta$ between 
successive bonds along chains 
are very small
($\langle \cos \theta \rangle \approx 1- \langle \theta^2 \rangle/2$), 
then Eqs.~(\ref{eq-Re2-dWLC}) and (\ref{eq-lp-dWLC}) become
\begin{equation}
     \langle R_e^2 \rangle = N\ell_b^2 \frac{1+\langle \cos \theta \rangle}
{1-\langle \cos \theta \rangle } \approx N \ell_b^2
\frac{4}{\langle \theta^2 \rangle} \,
\label{eq-Re2-FJC}
\end{equation}
and 
\begin{equation}
  \ell_p/\ell_b = 2 / \langle \theta^2 \rangle
\end{equation}
Eq.~(\ref{eq-Re2-FJC}) is equivalent to the mean square end-to-end distance 
of a freely jointed chain that $n_k$ Kuhn segments of length $\ell_K$ 
are jointed together,
\begin{equation}
   \langle R_e^2 \rangle = n_k \ell_k^2 = 2 \ell_p L \,.
\label{eq-Re2-FJC1}
\end{equation}
$L=N\ell_b=n_k \ell_k$ being the contour length and $\ell_K = 2 \ell_p$ in this limit.

In the continuum limit $\ell_b \rightarrow 0$, $N\rightarrow 0$, but
keeping $L$ and $\ell_p$ finite, we obtain from Eq.~(\ref{eq-Re2-dWLC})
the prediction for a continuous worm-like chain,
\begin{equation}
  \langle R_e^2 \rangle = 2 \ell_p L \left\{
1- \frac{\ell_p}{L} [1-\exp(-L/\ell_p) ] \right\} \,.
\label{eq-Re2-WLC}
\end{equation}
It gives the same result as that derived directly from the 
Kratky-Porod model~\cite{Kratky1949,Saito1967} for worm-like chains in $d=3$,
\begin{equation}
    {\cal H} =\frac{\ell_p k_BT}{2}\int_0^L \left(\frac{\partial^2 \vec{r}(s)}
{\partial s^2} \right)^2 ds \,,
\end{equation}
where the polymer chain is described by the contour $\vec{r}(s)$ in
continuous space.
Equation (\ref{eq-Re2-WLC}) describes the crossover behavior from
a rigid-rod for $L<\ell_p$, where $\langle R_e^2 \rangle =L^2$,
to a Gaussian coil for $L\gg \ell_p$, where $\langle R_e^2 \rangle =2 \ell_pL$
as shown in Eq.~(\ref{eq-Re2-FJC1}).

For semiflexible Gaussian chains the contour length $L=N\ell_b$ can also be
written as $L=n_p\ell_p$ and the mean square end-to-end distance and gyration radius
described in terms of $n_p$ and $\ell_p$ are~\cite{Kratky1949, Benoit1953}
\begin{equation}
 \frac{\langle R_e^2 \rangle}{2 \ell_p L} = 1-\frac{1}{n_p}[1-\exp(-n_p)] \,,
\label{eq-Re2np-WLC}
\end{equation}
\begin{equation}
\frac{6 \langle R_g^2 \rangle}{2 \ell_p L} = 1- \frac{3}{n_p}+\frac{6}{n_p^2}-\frac{6}{n_p^3}
[1-\exp(-n_p)] \, .
\label{eq-Rg2np-WLC}
\end{equation}
One can clearly recognize that Gaussian behavior of the radii is only
seen, if the number $n_p$ of the persistence length that fits to a given
contour length is large, $n_p \gg 1$, while a crossover to rigid-rod behavior
occurs for $n_p$ of order unity.
  
In recent works in Ref.~\cite{Wittmer2007,Hsu2010,Hsu2010b}, authors have 
shown that the exponential decay of the bond-bond 
orientational correlation function,
Eq.~(\ref{eq-cos-dWLC1}), and the Gaussian coil behavior, 
Eq.~(\ref{eq-Re2-WLC}) for $L\gg \ell_p$, predicted by
the worm-like chain model only hold for $s$ and $N$ up to some values $s^*$
and $N^*$, respectively when excluded volume 
effects are considered. The predictions of a theory based on the
Flory-type free energy minimization 
arguments~\cite{deGennes1979, Grosberg1994, Schaefer1980, Netz2003}
proposed as 
an alternative to semiflexible chains with excluded volume interactions
have been verified. In this treatment one considers a model where rods
of length $\ell_k$ and diameter $D$ are jointed together,
such that the contour length $L=N\ell_b=n_k \ell_k$. Apart from
prefactors of order unity, the second virial coefficient in $d=3$ then can
be estimated as 
\begin{equation}
    v_2 = \ell_k^2 D \,  .
\end{equation}
The free energy of a chain now contains two terms, 
the elastic energy taken
as that of a free Gaussian, i.e., $F_{el} \approx R_e^2 /(\ell_kL)$,
and the repulsive energy due to interactions
treated in mean field approximation, i.e. proportional to the square of
the density $n/R^3$ and the volume $R^3$. Hence,
\begin{equation}
   \Delta F /k_BT \approx R_e^2 /(\ell_K L) + v_2 R_e^3 [(L/\ell_K)/R_e^3]^2
\label{eq-F-Flory}
\end{equation}
Minimizing $\Delta F$ with respect to $R_e$, we obtain for 
$L\rightarrow \infty$ the standard Flory result 
\begin{equation}
   R_e \approx (v_1/\ell_k)^{1/5} L^{3/5} = (\ell_kD)^{1/5}(N\ell_b)^{3/5} \, .
\label{eq-R-Flory}
\end{equation}
Eq.~(\ref{eq-R-Flory}) holds also
for finite $L$ and $N>N^*=\ell_k^3/(\ell_bD^2)$ since the contribution of 
the second term in Eq.~(\ref{eq-F-Flory}) is still important.
For $N<N^*$ the first term in Eq.~(\ref{eq-F-Flory}) dominates, and the chain
behaves as a Gaussian coil,
$R_e^2=\ell_k L = \ell_k \ell_b N$, while for even smaller $N$,
$N<N^{\rm rod}=\ell_k/\ell_b$,
the chain behaves as a rigid-rod. Thus, the double crossover 
behavior of the mean square end-to-end distance is summarized as 
follows,
\begin{equation}
   \langle R_e^2 \rangle  \approx L^2\,, \quad N<N^{\rm rod}=\ell_k/\ell_b \quad 
{\rm (rod-like \, chain)}\,,
\end{equation}
\begin{equation}
   \langle R_e^2 \rangle \approx \ell_k L \,, \quad N^{\rm rod} < N < N^*
\quad {\rm (Gaussian \,coil)}\, ,
\end{equation}
\begin{equation}
    \langle R_e^2 \rangle \approx (\ell_k D)^{2/5} L^{6/5} \,, \quad  N>N^* 
\quad {\rm (SAW)}
\end{equation}

\begin{figure*}[ht]
\begin{center}
(a)\includegraphics[scale=0.29,angle=270]{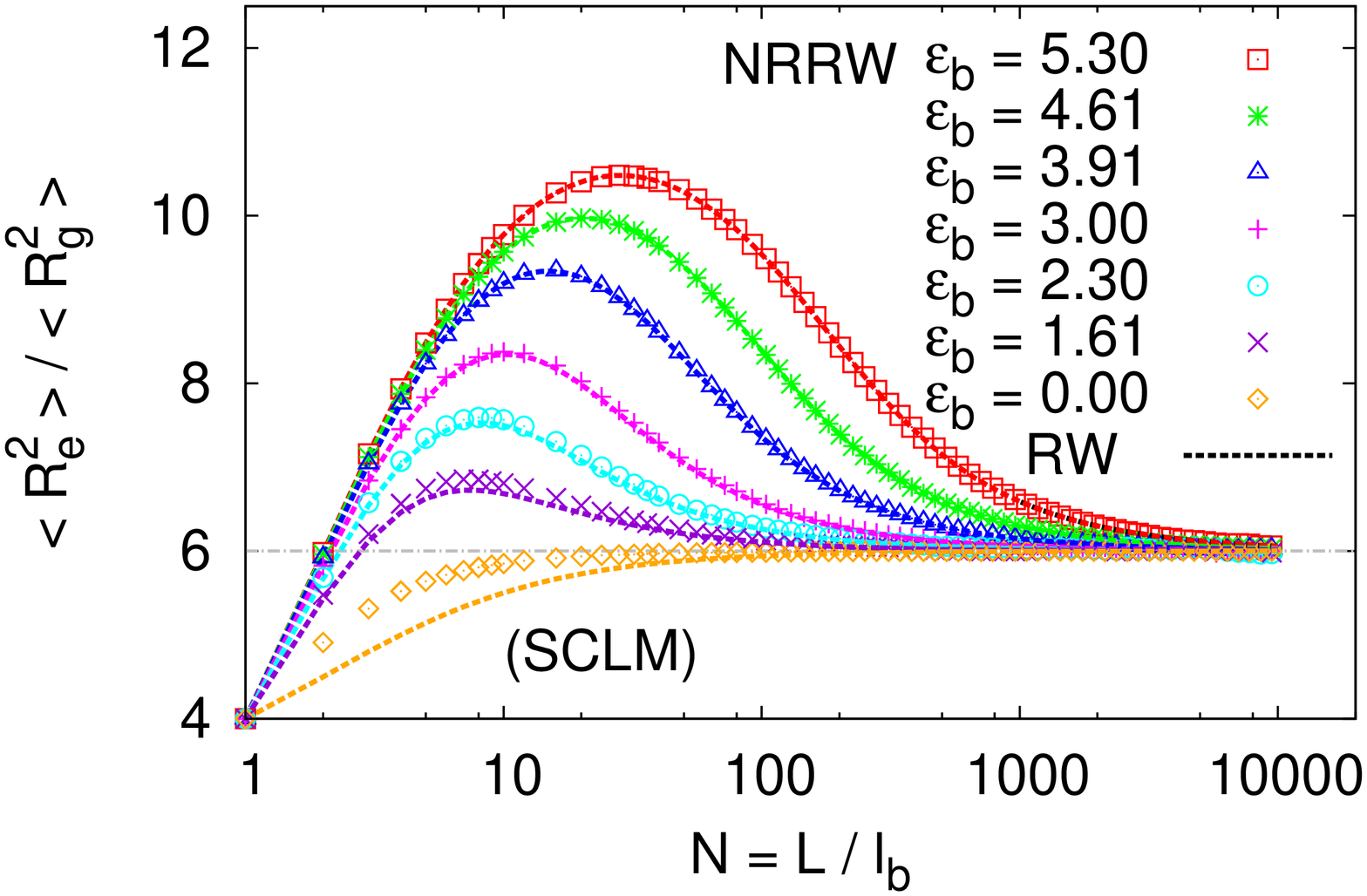} \hspace{0.4cm}
(b)\includegraphics[scale=0.29,angle=270]{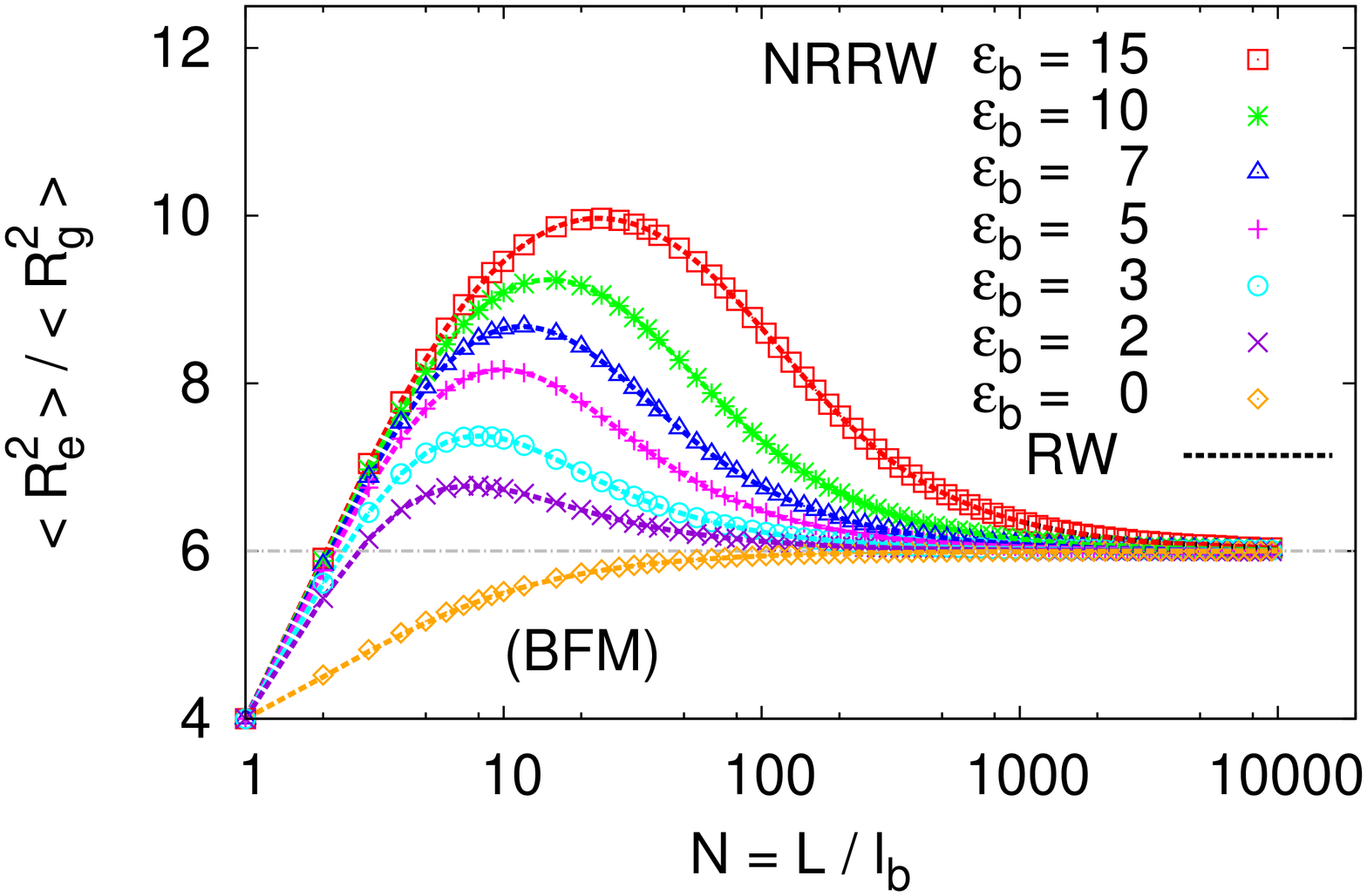}
\caption{Ratio between mean square end-to-end distance and
mean square gyration radius, $\langle R_e^2 \rangle/\langle R_g^2 \rangle$,
plotted against chain lengths (segments) $N=L/\ell_b$ for SCLM (a) and for BFM (b).
Data for semiflexible chains described by NRRWs and RWs are
shown by symbols and lines, respectively.
Here $\ell_b=1$ for SCLM, and $\ell_b=2.72$ for BFM.}
\label{fig-Reg2-semi}
\end{center}
\end{figure*}

\begin{figure*}[ht]
\begin{center}
(a)\includegraphics[scale=0.29,angle=270]{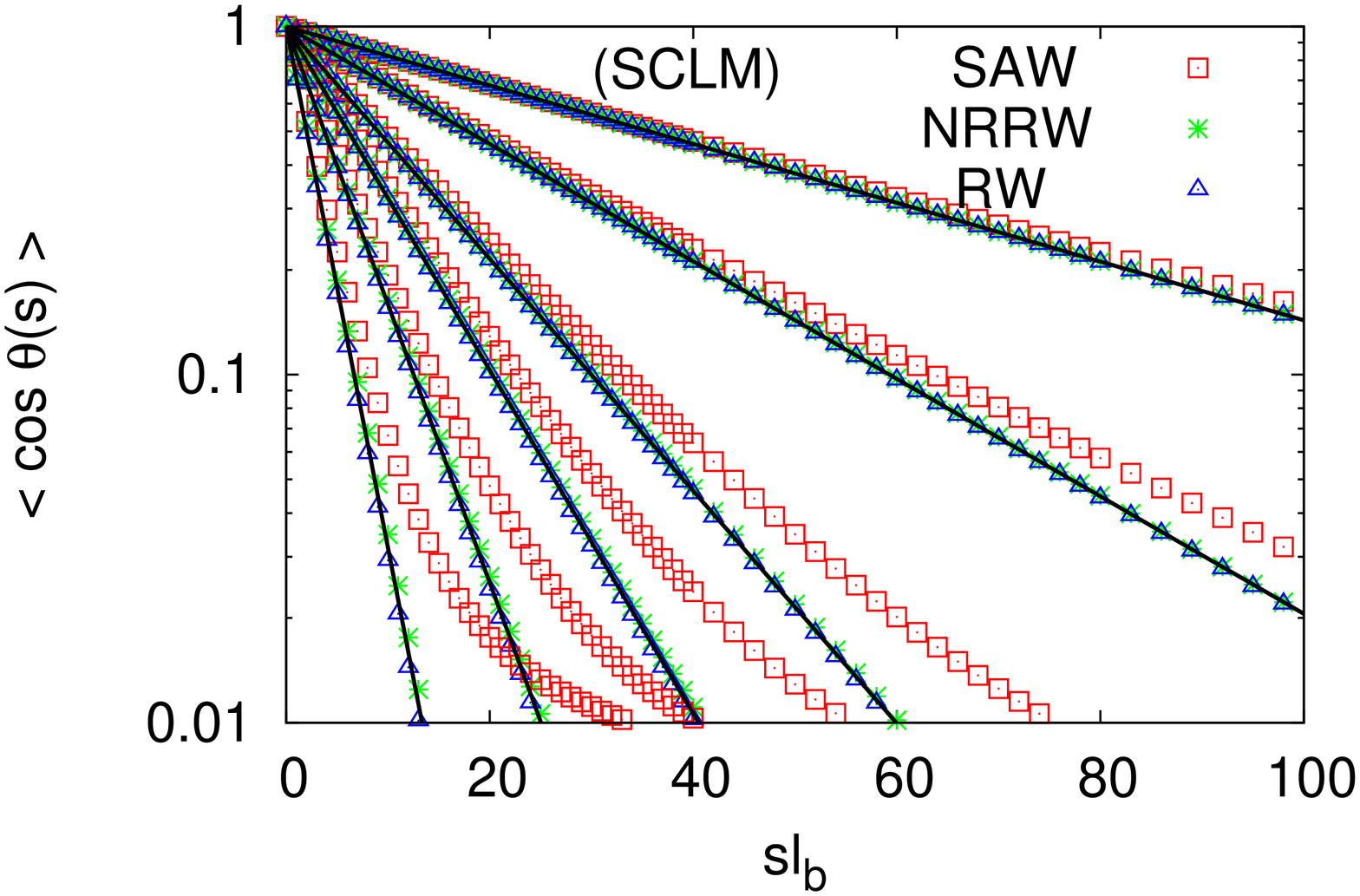} \hspace{0.4cm}
(b)\includegraphics[scale=0.29,angle=270]{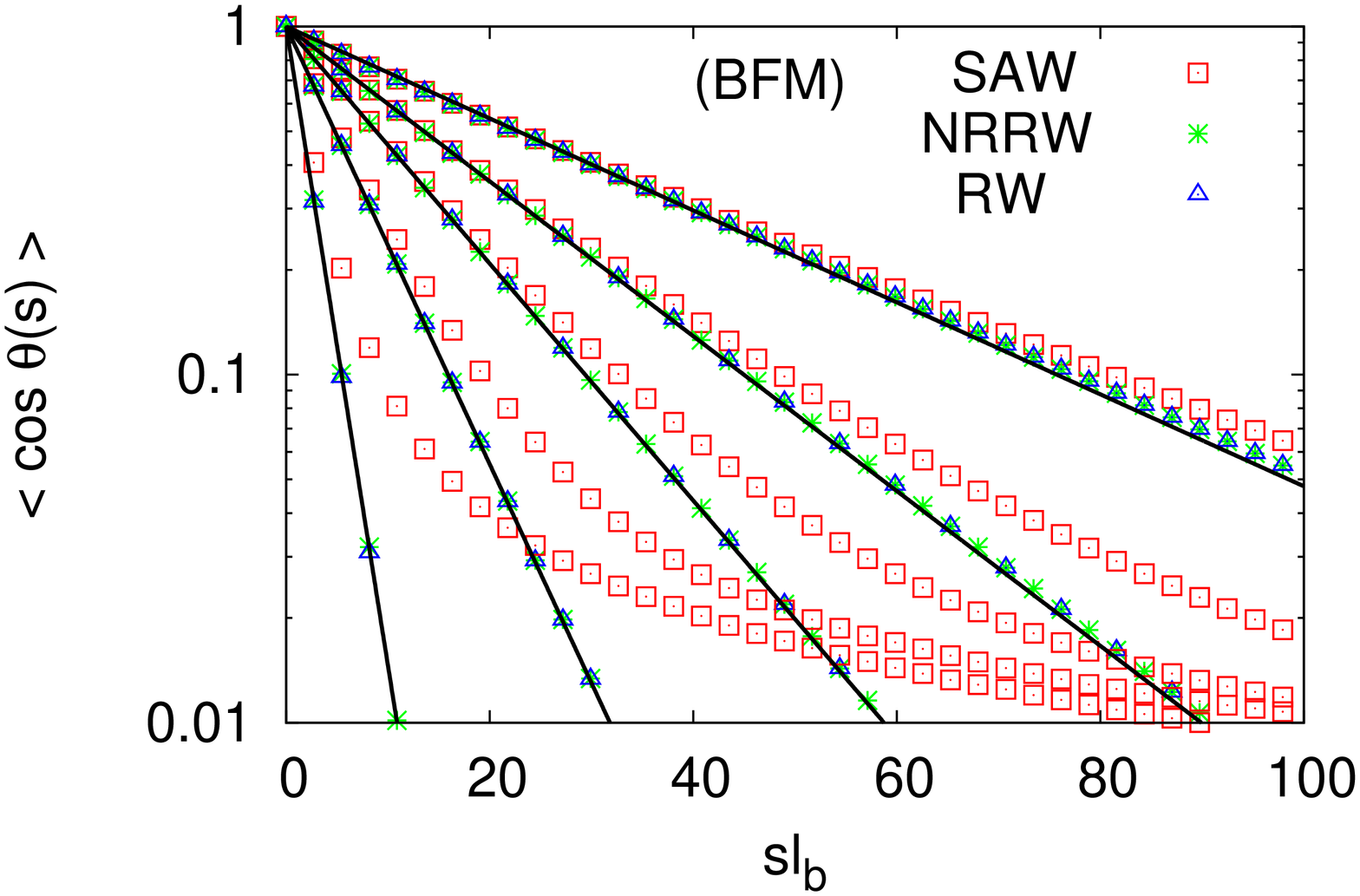}
\caption{Semi-log plot of the bond-bond orientational correlation function
$\langle \cos \theta (s) \rangle$ vs.\ $s\ell_b$ for SCLM with $\ell_b=1$
(a) and for BFM with $\ell_b=2.72$ (b).
Data are for semiflexible chains described by SAWs, NRRWs and RWs and 
for $\varepsilon_b=5.30$, $4.61$, $3.91$, $3.51$, $3.00$, $2.30$
from top to bottom in (a), and
for $\varepsilon_b=10$, $7$, $5$, $3$, and $1$ from top to bottom in (b). 
The straight lines
indicate fits of the initial decay, 
$\langle \cos \theta(s) \rangle \propto \exp(-s\ell_b/\ell_p)$ [Eq.~(\ref{eq-cos-dWLC1})],
for RWs.}
\label{fig-hs-semi}
\end{center}
\end{figure*}

\subsection{Simulation Results}

  In order to investigate the scaling behavior of the ratio $\langle R_e^2 \rangle /\langle R_g^2 \rangle$
for semiflexible RWs and NRRWs we plot our data $\langle R_e^2 \rangle /\langle R_g^2 \rangle$ versus
$N=L/\ell_b$ for several choices of the stiffness parameter (Fig.~\ref{fig-Reg2-semi}).
As $N$ increases, the data increase towards a maximum and then decrease towards
a plateau where the prediction
$\lim_{N \rightarrow \infty}\langle R_e^2 \rangle /\langle R_g^2 \rangle \approx 6$ for ideal chains holds.
At the location of the maximum of
$\langle R_e^2 \rangle /\langle R_g^2 \rangle$, $N=N_h$, the corresponding maximum
values is $h={\rm max} \; \langle R_e^2 \rangle /\langle R_g^2 \rangle$.
The maximum move monotonously to larger values as chains become stiffer.
The deviation between the data for RWs and NRRWs based on the SCLM decreases
as the bending energy $\varepsilon_b$ increases (Fig.~\ref{fig-Reg2-semi}a), while it is negligible
for the simulation data obtained based on the BFM (Fig.~\ref{fig-Reg2-semi}b) in all cases.

\begin{figure}[t]
\begin{center}
\includegraphics[scale=0.29,angle=270]{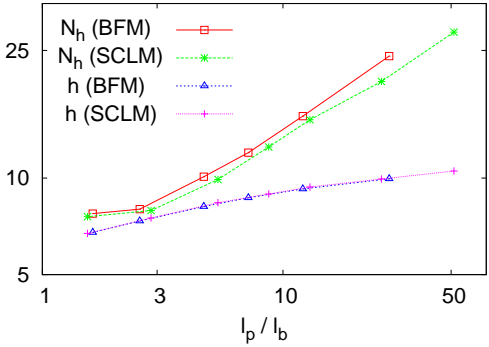} \hspace{0.4cm}
\caption{Location {$N_h$} and height $h$ of the maximum of
$\langle R_e^2 \rangle/\langle R_g^2 \rangle$ (see Fig.~\ref{fig-Reg2-semi})
plotted versus the 
persistence length $\ell_p/\ell_b$ for RWs based on
the SCLM and BFM.}
\label{fig-rmh-semi}
\end{center}
\end{figure}

  Figure~\ref{fig-hs-semi} shows the bond-bond orientational correlation 
function $\langle \cos \theta(s) \rangle$ plotted versus the chemical 
distance $s\ell_b$ covering
the range from flexible chains to stiff chains characterized by $\varepsilon_b$ for
the models SCLM and BFM. We compare the data obtained for SAWs, NRRWs, and RWs
for various values of $\varepsilon_b$. The intrinsic stiffness remains the same for
SAWs, NRRWs, and RWs as $\varepsilon_b$ is fixed.
Results obtained from both models verify that 
the asymptotic exponential decay of $\langle \cos \theta(s) \rangle$ is valid
only if the excluded volume effect is neglected, i.e., for RWs and NRRWs.
For semiflexible SAWs $\langle \cos \theta(s) \rangle \sim \exp(-s\ell_b/\ell_p)$
cannot be correct for $N\rightarrow \infty$~\cite{Wittmer2007, Hsu2010b},
we rather have 
\begin{equation}
         \langle \vec{b}_i \cdot \vec{b}_{i+s} \rangle \approx s^{-\beta}\, , \enspace
\beta=2-2\nu \approx 0.824\,, \enspace s^* \ll s \ll N \, .
\end{equation}
As we have seen in Fig.~\ref{fig-hs-semi}, the exponential decay is ill-defined for rather flexible 
SAWs. Using Eq.~(\ref{eq-lp-dWLC}) as a definition of the persistence length we
can still give the estimate of the persistence length
$\ell_{p,\theta}=-\ell_b/\ln \left[ \langle \cos \theta(s=1) \rangle \right]$
which is approximately the same as the estimate of the decay length $\ell_p$ for moderately stiff 
chains and stiff chains. 
The estimates of $\ell_p/\ell_b$ and $\ell_{p,\theta}/\ell_b$ depending on 
$\varepsilon_b$
using Eqs.~(\ref{eq-cos-dWLC1}) and (\ref{eq-lp-dWLC}) are listed in Table~\ref{table3} and ~\ref{table4}.
RWs are more flexible than NRRWs, and NRRWs are more flexible than
SAWs from the estimates of the persistence lengths $\ell_p/\ell_b$ and 
$\ell_{p,\theta}/\ell_b$ based 
on the SCLM. Using the BFM the persistence lengths are almost the same in all cases of $\varepsilon_b$
for RWs and NRRWs, and they are smaller compared with the estimates for SAWs.
Note that in Fig.~\ref{fig-hs-semi}b data deviate slightly from the 
fitting straight lines describing the
initial exponential decay for RWs and NRRWs as
the bending energy $\varepsilon_b$ increases, i.e., the stiffness of
chains increases. 
For $\varepsilon_b>10$ the problem is more severe. Therefore,
one should be careful using the BFM for studying rather stiff chains. 
An alternative way to the determination
of the persistence length would be given by the best fit of
the mean square end-to-end distance $\langle R_e^2 \rangle$ of RWs or NRRWs to Eq.~(\ref{eq-Re2-WLC}).
A simple exponential decay is always found for the probability distribution of
connected straight segments for semiflexible chains based on the SCLM~\cite{Hsu2010},
while large fluctuations are observed for semiflexible chains based on the BFM 
due to bond vector fluctuations and lattice artifacts~\cite{Wittmer1992}.
This is the main reason why the different scenarios of the bond-bond
orientational correlation functions between the SCLM and the BFM 
for stiff chains are seen in Fig.~\ref{fig-hs-semi}.
Figure~\ref{fig-rmh-semi} shows the locations
$N_h$ and the heights $h$
of the maximum of $\langle R_e^2 \rangle/\langle R_g^2 \rangle$
(Fig.~\ref{fig-Reg2-semi}) plotted versus the persistence length
$\ell_p/\ell_b$ for semiflexible RWs based on the SCLM and BFM.
Note that $N_h$, $h$, and $\ell_p/\ell_b$ all depend on $\varepsilon_b$ which
controls the stiffness of chains.
We see that the dependence between $h$ and $\ell_p/\ell_b$ are
the same for both models, while $N_h$ for the BFM is slightly larger than
that for the SCLM for a fixed value of $\ell_p/\ell_b$ since
chains based on the BFM are more flexible.

\begin{table*}[t]
\caption{
Two estimates for the persistence length $\ell_p/\ell_b$ from Eq.~(\ref{eq-cos-dWLC1})
and $\ell_{p,\theta}/\ell_b$ from Eq.~(\ref{eq-lp-dWLC}) for semiflexible RWs, NRRWs,
and SAWs with various values of $q_b(=\exp(-\varepsilon_b/k_BT))$ based
on the SCLM ($\ell_b=1$). Here in our simulations values of $q_b$ are chosen for convenience.}
\label{table3}
\begin{center}
\begin{tabular}{|l|c|ccccccccc|}
\hline
&      $q_b$ & 1.0 & 0.4 & 0.2 & 0.1 & 0.05 & 0.03 & 0.02 & 0.01 & 0.005 \\
&      ${\varepsilon_b}$ & 0.0 & 0.91 & 1.61 & 2.30 & 3.00 & 3.51 & 3.91 & 4.61 & 5.30  \\
\hline
$\ell_p/\ell_b$ & RW   & $\ldots$ &     0.84 & 1.54 & 2.83 & 5.37 & 8.73 & 12.95 & 25.67 & 51.38 \\
                & NRRW & $\ldots$ &     1.05 & 1.70 & 2.97 & 5.50 & 8.87 & 13.09 & 25.80 & 51.53 \\
                & SAW  & $\ldots$ & $\ldots$ & 2.04 & 3.35 & 5.96 & 9.54 & 13.93 & 26.87 & 52.61 \\
\hline
$\ell_{p,\theta}/\ell_b$ & RW & $\ldots$ & 0.84 & 1.54 & 2.83 & 5.36 & 8.73 & 12.95 & 25.66 & 51.37 \\
                         & NRRW & 0.62   & 1.05 & 1.70 & 2.98 & 5.50 & 8.87 & 13.08 & 25.79 & 51.53 \\
                         & SAW  & 0.67   & 1.12 & 1.81 & 3.12 & 5.70 & 9.10 & 13.35 & 26.28 & 51.52 \\
\hline
\end{tabular}
\end{center}
\end{table*}

\begin{table*}[htb]
\caption{
Two estimates for the persistence length $\ell_p/\ell_b$ from Eq.~(\ref{eq-cos-dWLC1})
and $\ell_{p,\theta}/\ell_b$ from Eq.~(\ref{eq-lp-dWLC}) for semiflexible RWs, NRRWs,
and SAWs with various values of $\varepsilon_b$ based
on the BFM ($\ell_b=2.72$).}
\label{table4}
\begin{center}
\begin{tabular}{|l|c|cccccccc|}
\hline
 & ${\varepsilon_b}$ &  0.0 & 1.0 & 2.0 & 3.0 & 5.0 & 7.0 & 10.0 & 15   \\
\hline
 $\ell_p/\ell_b$ & RW & $\ldots$ & 0.87 & 1.62 & 2.54 & 4.69 & 7.18 & 12.09 & 27.65 \\
                 & NRRW & $\ldots$ & 0.87 & 1.62 & 2.54 & 4.69 & 7.18 & 12.09 & 27.65 \\
                 & SAW & $\ldots$ & $\ldots$ & 1.91 & 2.78 & 4.94 & 7.39 & 12.37 & 27.93 \\
\hline
 $\ell_{p,\theta}/\ell_b$ & RW & $\ldots$ & 0.87 & 1.62 & 2.54 & 4.63 & 6.87 & 10.50 & 17.73 \\
                          & NRRW & 0.21 & 0.87 & 1.62 & 2.54 & 4.63 & 6.87 & 10.50 & 17.73 \\
                          & SAW & 0.61 & 1.11 & 1.80 & 2.65 & 4.68 & 6.90 & 10.52 & 17.75 \\
\hline
\end{tabular}
\end{center}
\end{table*}

\begin{figure*}[htb]
\begin{center}
(a)\includegraphics[scale=0.29,angle=270]{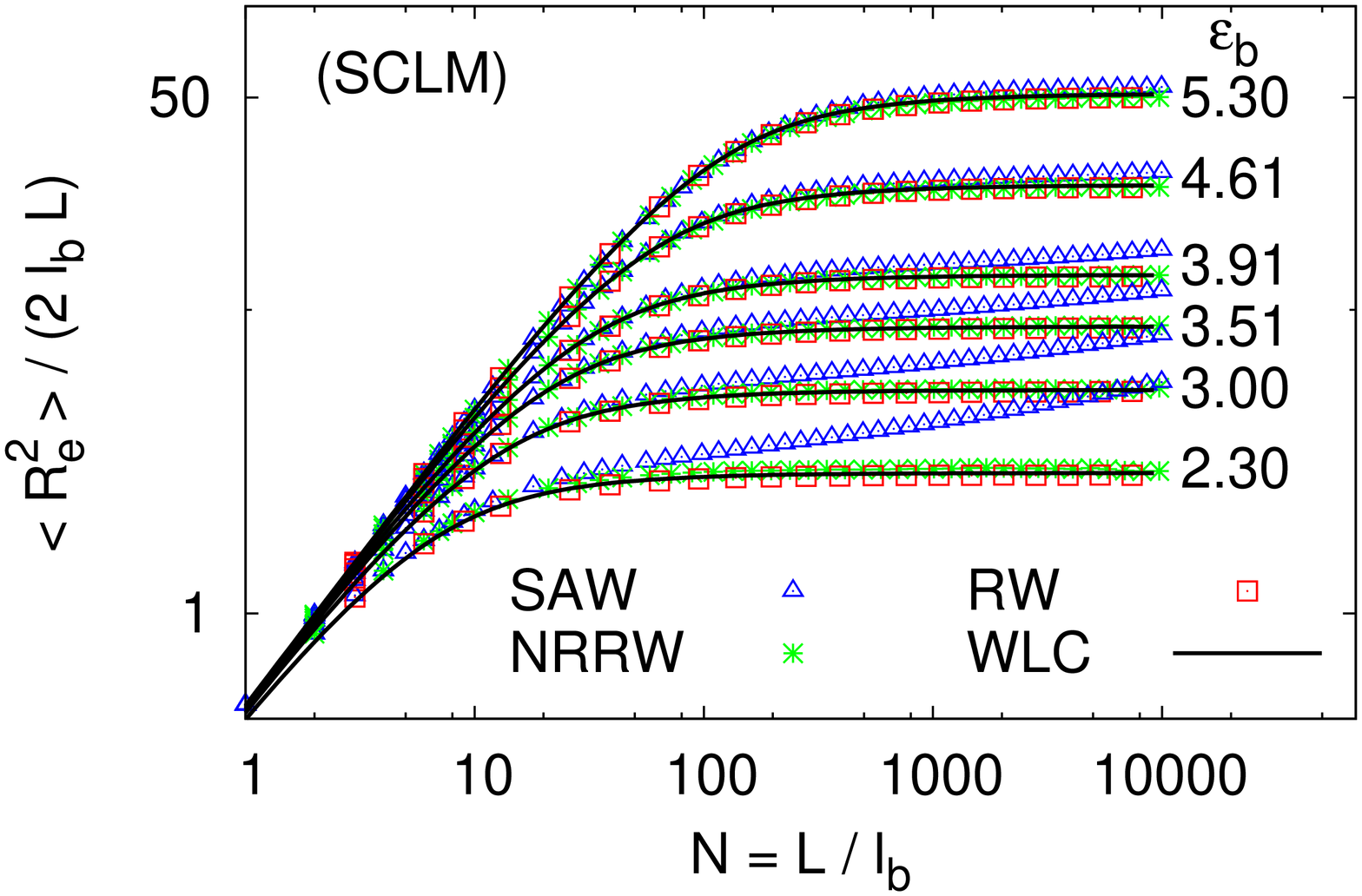}\hspace{0.4cm}
(b)\includegraphics[scale=0.29,angle=270]{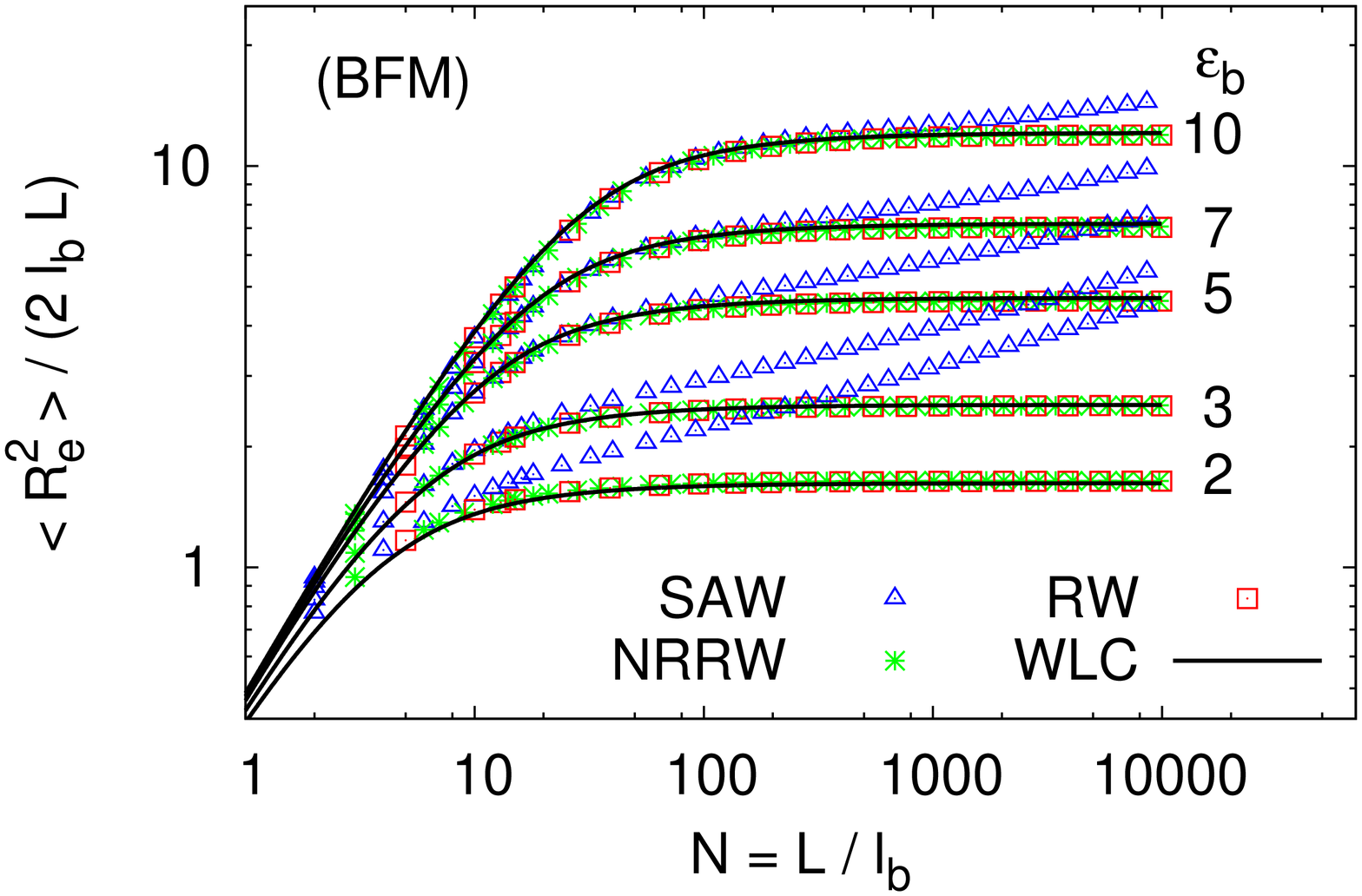}\\
(c)\includegraphics[scale=0.29,angle=270]{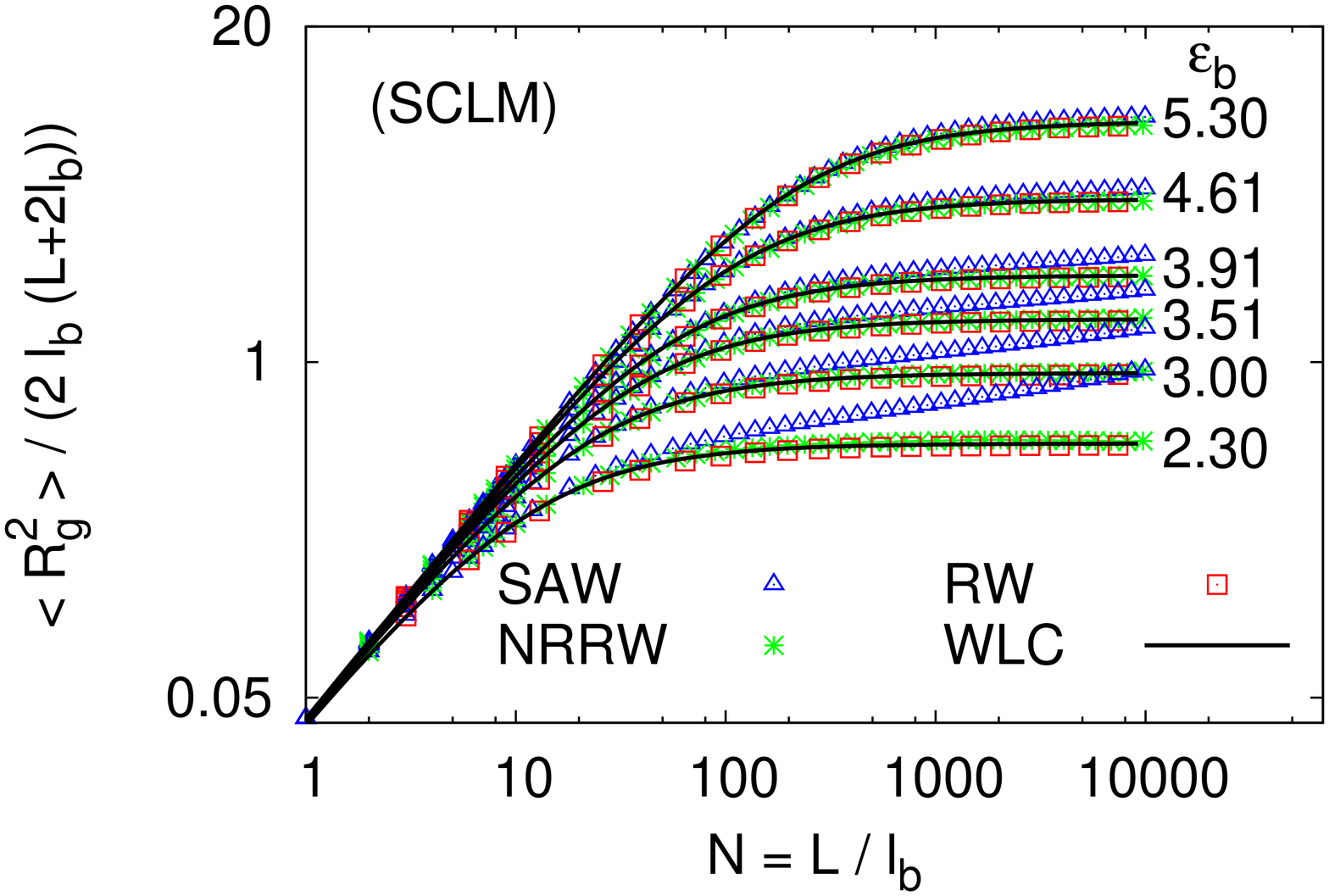}\hspace{0.4cm}
(d)\includegraphics[scale=0.29,angle=270]{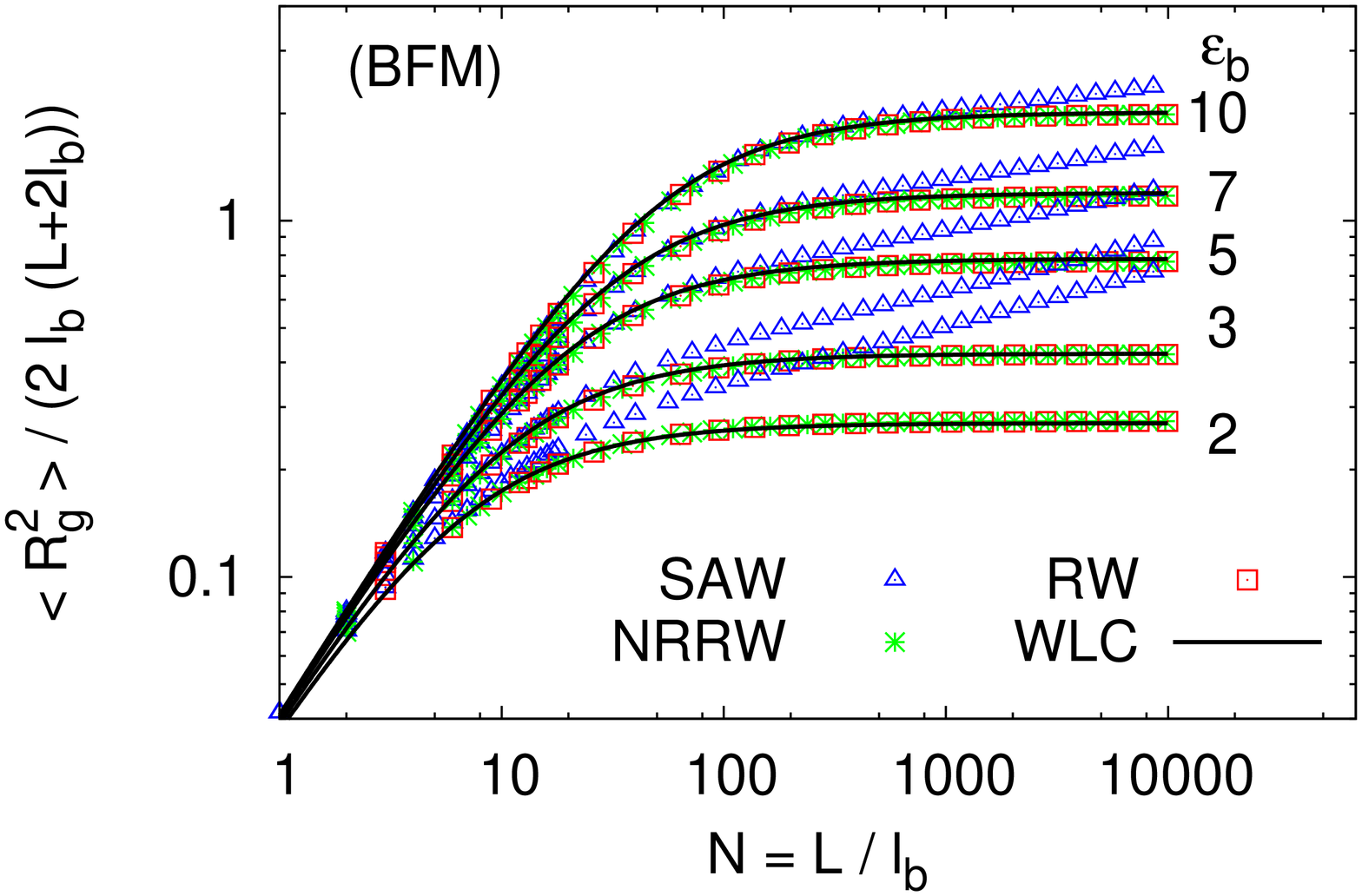}\\
\caption{Log-log plots of rescaled mean square end-to-end distance
$\langle R_e^2 \rangle /(2 \ell_b L)$ (a),(b) and
rescaled mean square gyration radius
$\langle R_g^2 \rangle /(2 \ell_b (L+2\ell_b))$ (c),(d)
versus ${N=L/\ell_b}$ for semiflexible chains described by
SAWs, NRRWs and RWs based on the SCLM with $\ell_b=1$ (a),(c) and BFM with
$\ell_b=2.72$ (b),(d).
Data for various values of $\varepsilon_b$ are shown, as indicated.
Solid curves refer to the theoretical prediction, Eq.~(\ref{eq-Re2-WLC}), for WLC.
The values of the persistence length $\ell_p/\ell_b$ for NRRW are taken
from Tables~\ref{table3} and \ref{table4}.
}
\label{fig-Re2-semi}
\end{center}
\end{figure*}

The scaling plots for testing the applicability of the worm-like chain prediction, 
Eq.~(\ref{eq-Re2-WLC}) and Eq.~(\ref{eq-Rg2np-WLC})
to our data of $\langle R_e^2 \rangle$ and $\langle R_g^2 \rangle$ 
are shown in Fig.~\ref{fig-Re2-semi}.
The persistence length $\ell_p/\ell_b$ in Eq.~(\ref{eq-Re2-WLC}) for various 
values of $\varepsilon_b$
are extracted from the exponential fit of Eq.~(\ref{eq-cos-dWLC1}) for NRRWs 
(see Tables~\ref{table3} and \ref{table4}).
Since the worm-like chain model is formulated in the continuum, care has to 
be taken to correctly take into account the lattice structure of the present 
model, particularly in the rod limit.
Assuming that a rigid rod consisting of $N$ monomers is located at 
$x_1=\ell_b$, $x_2=2\ell_b$, 
$\ldots$, $x_N=N$ along the x-axis on the simple cubic lattice, the mean square 
gyration radius is:
\begin{eqnarray}
    \langle R_e^2 \rangle_{\rm rod} &=& \frac{1}{N}\sum_{k=1}^N (k \ell_b)^2 -
\left( \frac{1}{N} \sum_{k=1}^N k\ell_b \right)^2  \nonumber \\
&=& \frac{(N+1)(2N+1)\ell_b^2}{6}-\frac{(N+1)^2\ell_b^2}{4} \nonumber \\
& =& \frac{(N+1)(N-1)\ell_b^2}{12}=\frac{L(L+2\ell_b)}{12} \,.
\label{eq-Rg2-WLC}
\end{eqnarray}
Therefore, due to the lattice structure,
the mean square gyration radius is rescaled by
{$(L+2\ell_b)$} instead of $L$ in order to compare with the 
theoretical predictions in Fig.~\ref{fig-Re2-semi}c,d.
For semiflexible RWs and NRRWs the data are indeed very well described by the 
worm-like chain model.
As $N$ increases, we observe the crossover behavior from a rigid-rod regime
to a Gaussian coil regime. The plateau value in the Gaussian regime corresponds to the
persistence length $\ell_p/\ell_b$ in Fig.~\ref{fig-Re2-semi}a,b and
$(1/6)\ell_p/\ell_b$ in Fig.~\ref{fig-Re2-semi}c,d.
For SAWs the deviation from the prediction becomes more prominent as chains are more flexible
since the excluded volume effects are more important.

\begin{figure*}[t]
\begin{center}
(a)\includegraphics[scale=0.29,angle=270]{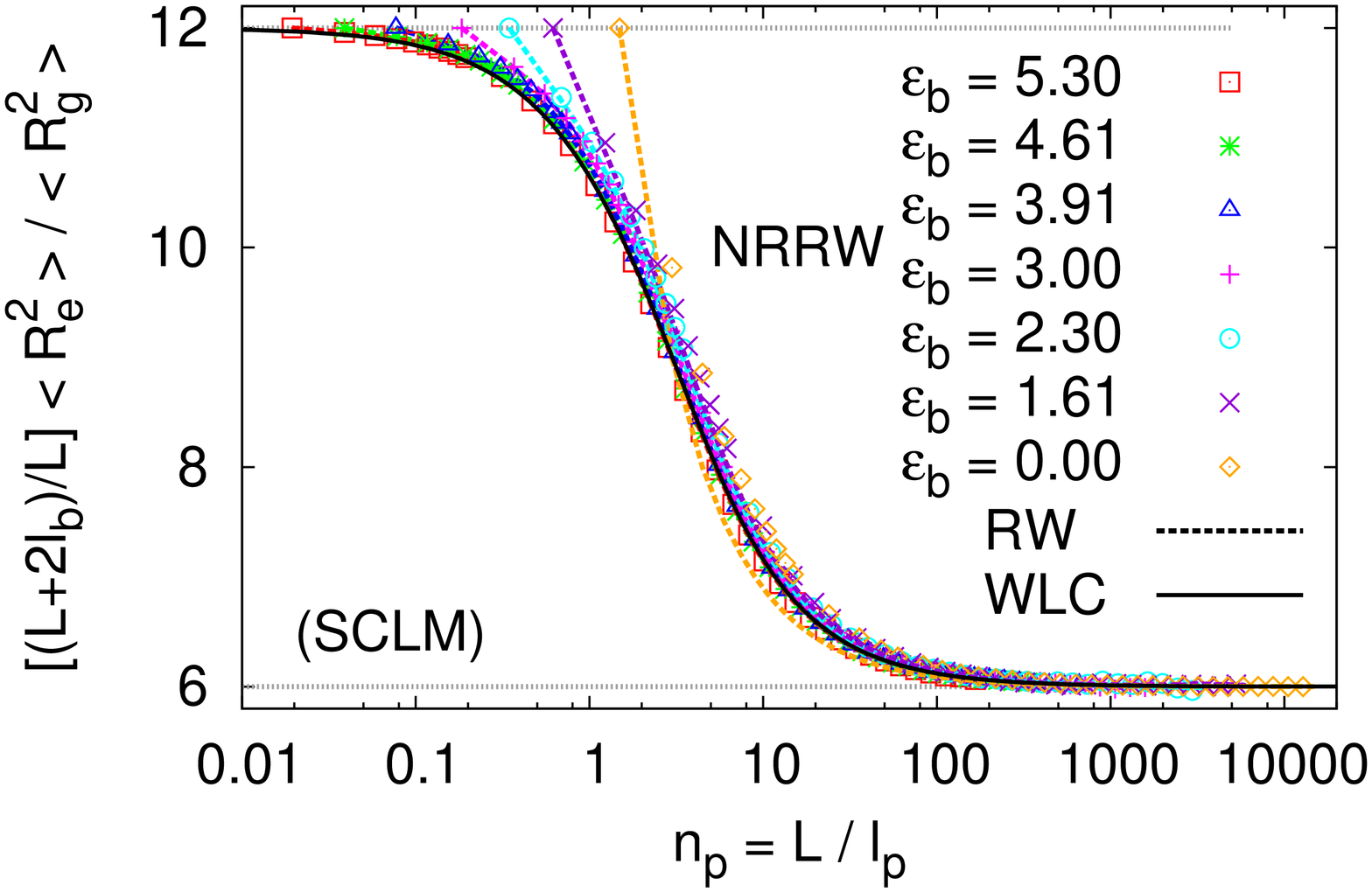}\hspace{0.4cm}
(b)\includegraphics[scale=0.29,angle=270]{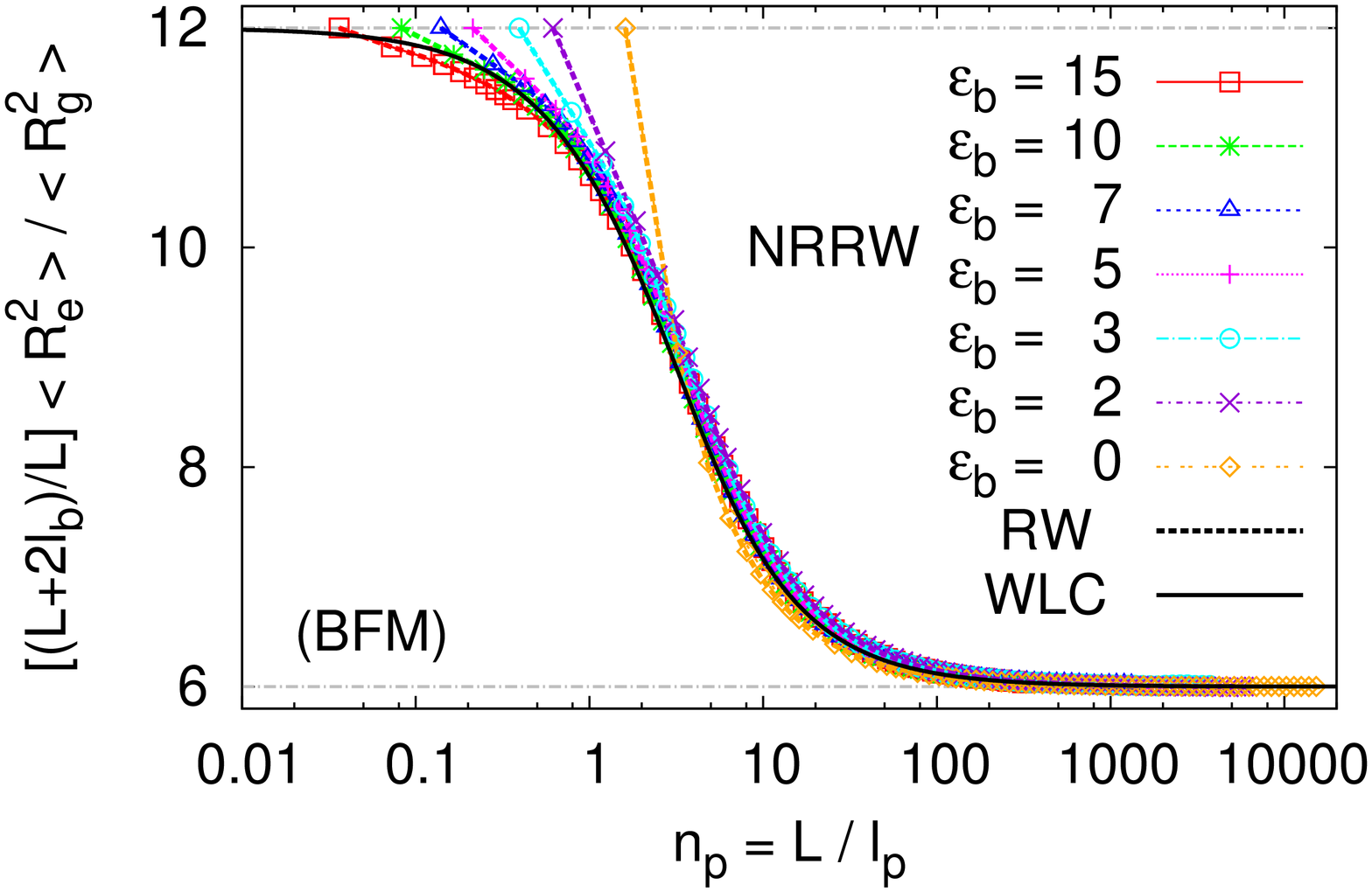}\\
(c)\includegraphics[scale=0.29,angle=270]{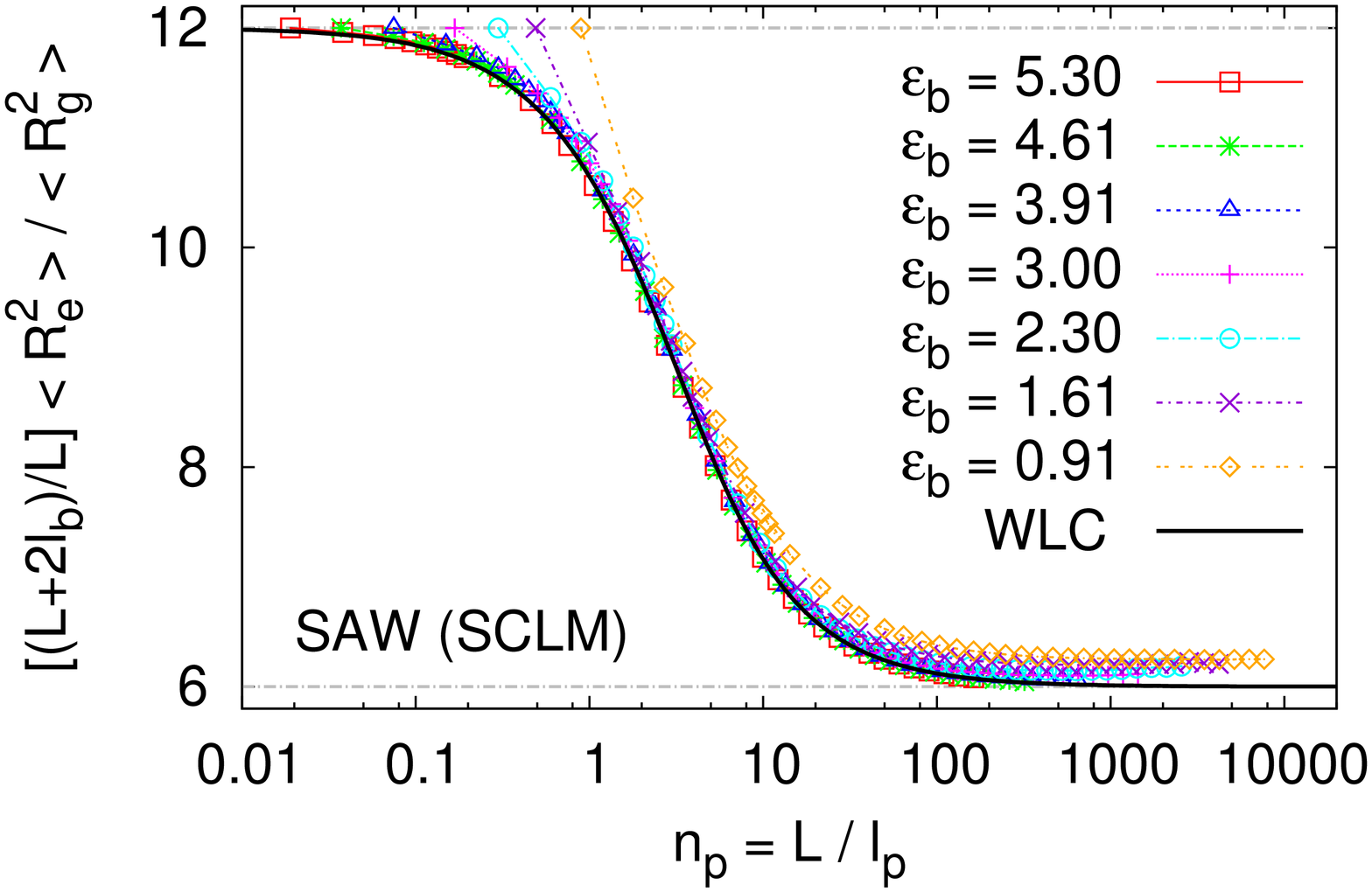}\hspace{0.4cm}
(d)\includegraphics[scale=0.29,angle=270]{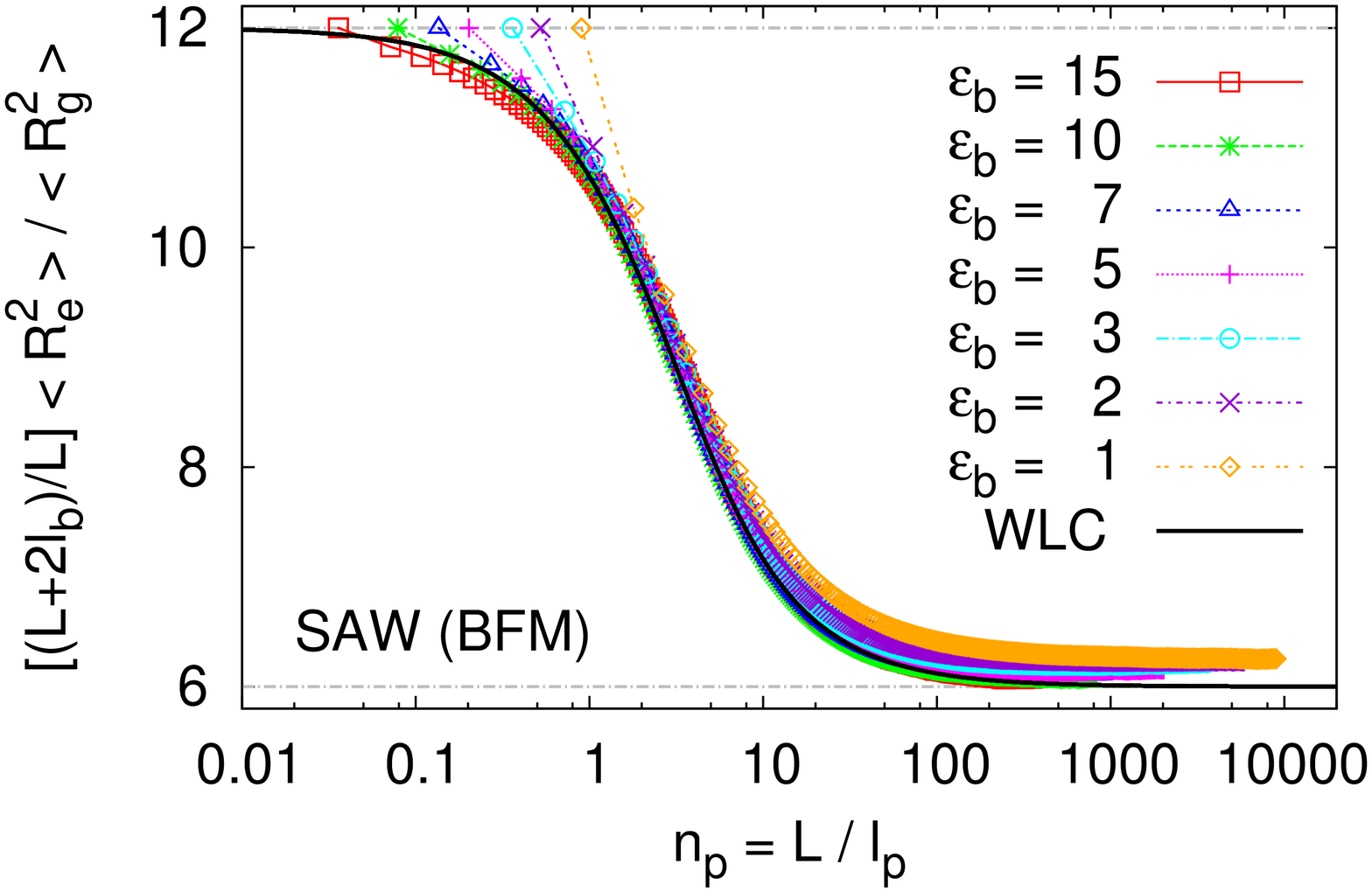}
\caption{Semi-log plots of $[(L+2\ell_b)/L]\langle R_e^2 \rangle /\langle R_g^2 \rangle$
versus $n_p=L/\ell_p$ for semiflexible chains described by RWs and NRRWs (a),(b)
and SAWs (c),(d).
Data for various values of $\varepsilon_b$ are shown, as indicated.
Solid curves refer to the theoretical prediction, the ratio between
Eq.~(\ref{eq-Re2np-WLC}) and Eq.~(\ref{eq-Rg2np-WLC}), for WLC.
The values of the persistence length $\ell_p/\ell_b$ for RWs, NRRWs, and SAWs are taken
from Tables~\ref{table3} and \ref{table4}, respectively.
}
\label{fig-Reg2np-semi}
\end{center}
\end{figure*}

Note that one should not consider the correction factor $(L+2\ell_b)/L$ 
relative to the Kratky-Porod model in Eq.~(\ref{eq-Rg2-WLC}) as a 
``lattice artefact": in a real stiff polymer (e.g.\ an alkane-type chain) 
one also has a sequence of discrete individual monomers (separated by almost 
rigid covalent bonds along the backbone of the chain) lined up linearly 
(like in a rigid rod-like molecule) over about the distance of a persistence 
length.
Furthermore, we compare simulation results of the ratio 
$\langle R_e^2 \rangle/\langle R_g^2 \rangle$ multiplied by
$[(L+2\ell_b)/L]$ as a function of ${n_p=L/\ell_p}$ to the theoretical prediction, 
the ratio between Eq.~(\ref{eq-Re2np-WLC}) and Eq.~(\ref{eq-Rg2np-WLC}),
in Fig.~\ref{fig-Reg2np-semi}. We see the nice data
collapse for RWs and NRRWs in the Gaussian regime ({$n_p \gg 1$}) and the increase of
deviations from the master curve
as the stiffness of chains decreases in Fig.~\ref{fig-Reg2np-semi}a,b.
The ratio $[(L+2\ell_b)/L]\langle R_e^2 \rangle/\langle R_g^2 \rangle \approx 12$ as 
${n_p \rightarrow 0}$ for a rigid-rod, 
while ${[(L+2\ell_b)/L]\langle R_e^2 \rangle/\langle R_g^2 \rangle \approx 6}$
as $n_p \rightarrow \infty$ for a Gaussian coil.
For SAWs we still see the nice data collapse in Fig.~\ref{fig-Reg2np-semi}c,d, but
in both rigid-rod and Gaussian coil regimes the deviations from the master curve
become more prominent as chains are more flexible. For ${n_p>1}$ the deviation
is due to the excluded volume effects, and finally
$[(L+2\ell_b)/L]\langle R_e^2 \rangle/\langle R_g^2 \rangle \approx 6.25$
as $n_p \rightarrow \infty$ for SAWs.
Note that in both models the ratio of the mean square end-to-end and gyration 
radii exceed its asymptotic value still significantly even if $n_p$ is as 
large as ${n_p = 10}$.

{Recently, Huang et al.~\cite{Huang2014a, Huang2014b} performed Brownian
dynamics simulations on two-dimensional (2D) semiflexible chains described
by a BSM including the excluded volume interactions. 
Varying the chain stiffness and chain length their results
confirmed the absence of a Gaussian regime in agreement with the results 
from semiflexible SAWs based on the SCLM~\cite{Hsu2011}, and with 
observations from experiments of circular single stranded DNA adsorbed 
on a modified graphite surface~\cite{Rechendorff2099}.
The rescaled mean square end-to-end distance, 
$\langle R_e^2 \rangle/(2L\ell_p)$,
in terms of $L/\ell_p$ for both models on the lattice and in the continuum
turns out to be universal from the rigid-rod regime up to the crossover 
regime ($L/\ell_p \sim 1$) irrespective
of the models chosen for the simulations. In the 2D SAW regime, 
different amplitude factors result from the different models~\cite{Huang2014c}}.

{
In $d=3$, we indeed see the nice data collapse for semiflexible 
RWs, NRRWs, and SAWs 
in the plot of $\langle R_e^2 \rangle /(2 L \ell_p)$ versus $L/2\ell_p$
(cf. Fig.~\ref{fig-Re2-semi}a,b) from rod-like regime 
crossover to the Gaussian regime for $N < N^*$ (not shown), 
and the data obtained from the two lattice
models are well described by the Kratky-Porod scaling function, 
Eq.~(\ref{eq-Re2-WLC}).
For the BSM in the continuum we should expect the same universal behavior.
Although for semiflexible SAWs the second crossover from the Gaussian
regime to the SAW regime for $N > N^*$ is rather gradual and not sharp, 
the relationship~\cite{Hsu2012} between the crossover chain length $N^*$
and the persistence length $\ell_p/\ell_b$, 
$N^* \propto (\ell_p/\ell_b)^{2.5}$, holds for these two models here.
It would be interesting to check whether such a scaling law would also 
hold for the BSM.}

\begin{figure*}[htb]
\begin{center}
(a)\includegraphics[scale=0.29,angle=270]{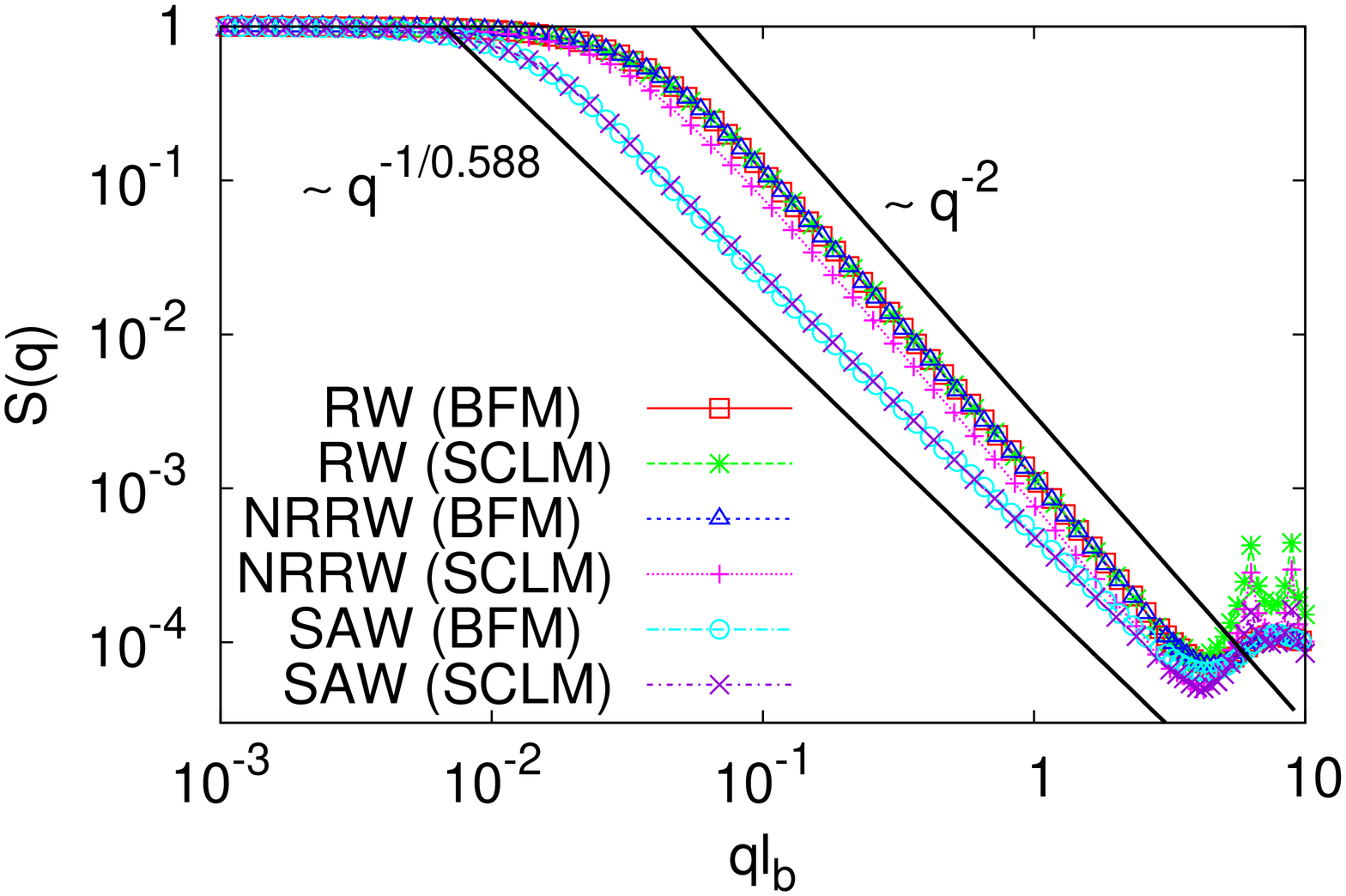}\hspace{0.4cm}
(b)\includegraphics[scale=0.29,angle=270]{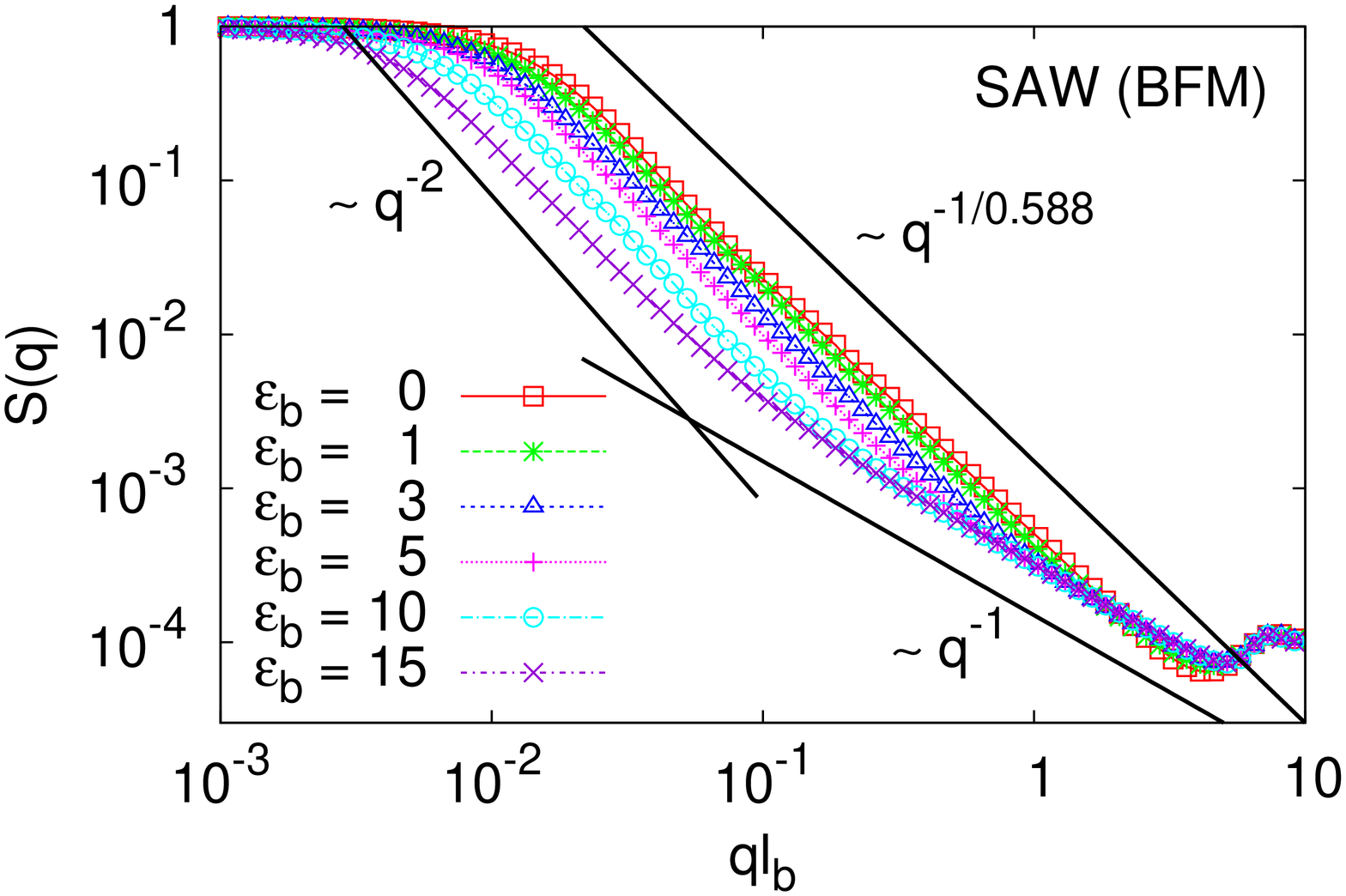}\\
(c)\includegraphics[scale=0.29,angle=270]{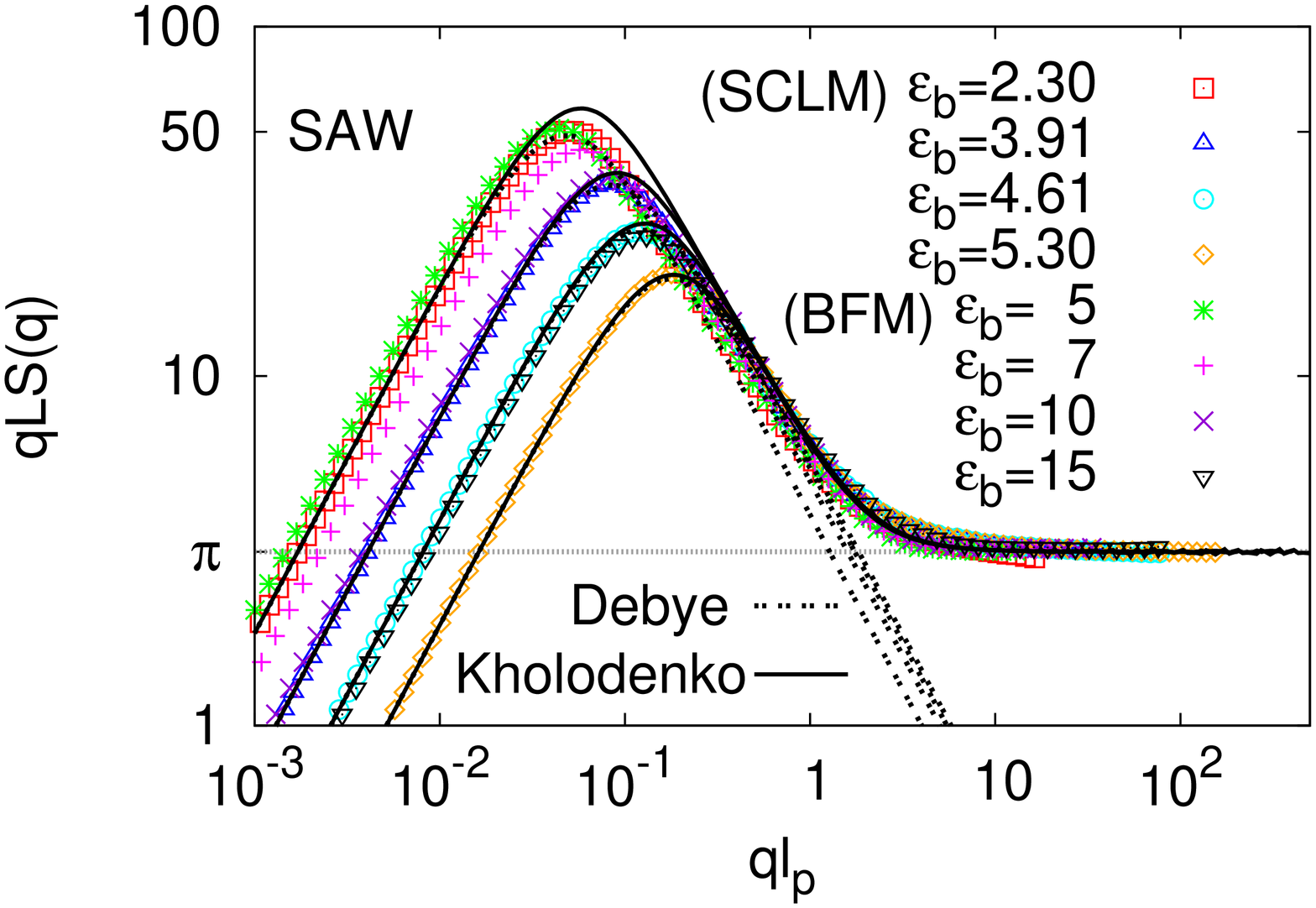}\hspace{0.4cm}
\caption{(a)(b) Log-log plot of structure factor $S(q)$ versus $q\ell_b$. Data are
for fully flexible chains based on the SCLM and BFM in (a), and for semiflexible chains
based on the BFM including $6$ choices of the stiffness in (b).
(c) Rescaled structure factor $qLS(q)$ plotted versus $ql_p$. Data are for semiflexible
chains based on the SCLM and BFM including $4$ choices of the stiffness each.
In (a)(b), the straight lines indicate the rod-like behavior at large $q$ (slope $=-1$)
the SAW behavior for flexible chains (slope $=-1/\nu$, with $\nu=0.588$), and
the Gaussian behavior (slope $=-2$). In (c), the formulas proposed by Kholodenko
\{Eqs.~(\ref{eq-Kholodenko1})-(\ref{eq-Kholodenko3})\},
the Debye function \{Eq.~(\ref{eq-Debye})\} for
Gaussian chains, and $qLS(q) \rightarrow \pi$ for a rigid-rod~\cite{Neugebauer1943} are also shown for
comparison.}
\label{fig-sq-semi}
\end{center}
\end{figure*}

{ 
The structure factor $S(q)$ is an experimentally accessible quantity
measured by neutron scattering. We therefore also estimate 
$S(q)$ by
\begin{equation}
     S(q)= \frac{1}{(N+1)^2} \left \langle \sum_{i=0}^N \sum_{j=0}^N 
\exp (i \vec{q} \cdot [\vec{r}_i-\vec{r}_j]) \right \rangle
\end{equation}
where $\{\vec{r}_i\}$ denote the positions of the $(N+1)$ monomers in a
chain, and the structure factor is normalized such that $S(q \rightarrow 0)=1$.
In order to compare the results of $S(q)$ obtained for fully flexible
RWs, NRRWs, and SAWs based on the SCLM and BFM, we plot $S(q)$ versus $q\ell_b$ 
($\ell_b=1$ for SCLM, and $\ell_b=2.72$ for BFM) in 
Fig.~\ref{fig-sq-semi}a. We see that 
$S(q) \approx 1-q^2 \langle R_g^2 \rangle/3$ for $q\rightarrow 0$,
while for $q \gg \sqrt{\langle R_g^2 \rangle}$ the power
law $S(q) \sim q^{-1/\nu}$ ($\nu=0.588$ for SAWs,
and $\nu=0.5$ for RWs and NRRWs) holds. The lattice artifact sets in at $q \ell_b \approx \pi$.
Due to the local packing the first peak appears at $q\ell_b \approx 2 \pi$ for the SCLM, while 
at $q \ell_b \approx 2.4 \pi$ for the BFM as $q$ increases.
In Fig.~\ref{fig-sq-semi}b we show the results for semiflexible SAWs of different stiffnesses
based on the BFM. The Gaussian regime where $S(q) \sim q^{-2}$ for large values of $\varepsilon_b$
and then crosses gradually over to
$S(q) \sim q^{-1}$ as expected for rigid rods.~\cite{Neugebauer1943}.}

{
Finally we analyze the structure factor $S(q)$ in the form of Kratky-plots, $qLS(q)$ plotted
versus $q\ell_p$, shown in Fig.~\ref{fig-sq-semi}c for semiflexible SAWs. 
Data are only for $q<\pi$.
The well-known theoretical predictions of the scattering from rigid-rods~\cite{Neugebauer1943},
$qLS(q) \rightarrow \pi$, and
Gaussian chains, the Debye function~\cite{deGennes1979,Cloizeaux1990,Schaefer1999,Higgins1994},
\begin{equation}\label{eq-Debye}
   S_{\rm Debye}(q) = \frac{2}{q^2 \langle R_g^2 \rangle}
\left\{ 1-\frac{1}{q^2 \langle R_g^2 \rangle} [1-\exp(-q^2 \langle R_g^2 \rangle)] \right\}\,,
\end{equation}
and the interpolation formula which describes the two limiting cases of
Gaussian coils and rigid rods exactly by Kholodenko~\cite{Kholodenko1993},
\begin{equation}\label{eq-Kholodenko1}
S(q) = \frac 2 x [I_1(x)-\frac 1 x I_2 (x)], \quad x = 3L/2 \ell_p
\end{equation}
where $I_n(x) = \int \limits _0^x dzz^{n-1}f(z)$,
and the function $f(z)$ is given by
\begin{eqnarray}\label{eq-Kholodenko2}
f(z) =  \left\{ \begin{array}{r@{\, , \quad}l}
\frac{1}{E} \frac{\sinh (Ez)}{\sinh z} & q \leq \frac{3}{2 \ell_p}  \, , \\
\frac {1} {\hat{E}} \frac {\sin (\hat{E}z)} {\sinh z} & q >\frac{3}{2\ell_p}\,,
\end{array} \right.
\end{eqnarray}
with
\begin{equation}\label{eq-Kholodenko3}
E = \left[1-\left(\frac {2q\ell_p}{3} \right)^2 \right]^{1/2}\, , \quad 
\hat{E} = \left[\left(\frac {2q\ell_p}{3}\right)^2 -1\right]^{1/2}\;,
\end{equation}
are also shown for comparison~\cite{Hsu2012b}.
Near the peak the discrepancy of our data from the theoretically predicted
formulas increases as the bending energy $\varepsilon_b$ decreases showing that
the excluded volume effect sets in. For semiflexible polymer chains of almost the same 
persistence length based on the two different lattice models, 
the structure factors are on top of each other.
}

\section{Conclusions}
\label{conclusion}
In this paper we have studied single polymer chains covering the
range from fully flexible chains to stiff chains under very good solvent
conditions by extensive Monte Carlo simulations based on two
coarse-grained lattice models: the standard simple cubic lattice model 
and the bond fluctuation model. With the pruned-enriched Rosenbluth method
the conformations of polymer chains mimicked by random walks, non-reversal
random walks, and self-avoiding walks depending on the effective interactions between
monomers have been analyzed in detail. 
We give the precise estimate of the fugacity $\mu_\infty$ and the entropic 
exponent $\gamma$ for self-avoiding walks based on the bond fluctuation model.
The universal scaling predictions of mean square end-to-end distance, 
$\langle R_e^2 \rangle$ [Eq.~(\ref{eq-Re})],
and mean square gyration radius, $\langle R_g^2 \rangle$ [Eq.~(\ref{eq-Rg})], for
fully flexible chains are verified as
one should expect, and the corresponding amplitudes $A_e$ and $A_g$ 
depending on the models are determined. We have also
checked the probability distributions of $R_e$ and $R_g$, $P(R_e)$ and $P(R_g)$,
respectively. Especially we point out that the previous
estimate of the parameter $A$ in Eqs.~(\ref{eq-A}) for SAWs is an overestimate due to
the finite-size effect. Our results also agree with the results
based on the BSM~\cite{Vettorel2010}, that the formula Eq.~(\ref{eq-pRg}) 
predicted by Lhuillier~\cite{Lhuillier1988} is a good approximate
formula for RWs.  

For semiflexible chains the additional regime of rod-like behavior
causes slow transients in many quantities, before the asymptotic behavior
of flexible chains is reached (see e.g. Fig.~\ref{fig-Reg2-semi}).
In the absence of the excluded volume effect, a single crossover 
occurs, from rigid-rods to Gaussian coils as implied by the Kratky-Porod model,
while a double crossover occurs from rigid-rods to Gaussian coils and then
to swelling coils due to the excluded volume interaction as predicted 
by the Flory-like arguments. 
We have verified the Kratky-Porod crossover scaling behavior for semiflexible RWs,
semiflexible NRRWs, 
and for semiflexible SAWs when the excluded volume effect is not yet important,
otherwise the Flory prediction takes over for semiflexible SAWs.
The flexibility of chains in our model is
controlled by the bending potential ${U_b=\varepsilon_b(1-\cos \theta)}$.
Our results of bond-bond orientational correlation functions 
${\langle \cos \theta(s) \rangle}$  (Fig.~\ref{fig-hs-semi})
show that the persistence lengths of semiflexible RWs, NRRWs, and SAWs are the
same for a given bending energy $\varepsilon_b$ based on the same lattice model.
But, with a caveat: there is a problem of fitting the exponential decay to 
the data of ${\langle \cos \theta(s) \rangle}$ for not only 
semiflexible SAWs but also semiflexible RWs and NRRWs 
based on the BFM for ${\varepsilon_b>10}$ (rather stiff chains) due to the fluctuations of bonds and
the lattice artifacts as it was mentioned in Ref.~\cite{Wittmer1992}.
{
The structure factor describing the scattering from semiflexible linear polymer chain
based on the SCLM provides an almost perfect match to the result based on the BFM
when we adjust the bending energy $\varepsilon_b$ such that the same persistence length 
$\ell_p$ results for both models.}

From our simulations the different crossovers to the asymptotic behavior 
of single chains based on the SCLM and BFM are observed and investigated.
Similar effects have to be expected for real chemical systems as well.
Thus coarse graining will require different mapping ratios for different
coarse-grained models.
However, the equilibration time may rise dramatically 
for simulating large and complex realistic polymer systems.
A proper mapping onto a coarse-grained model where the
number of degrees of freedom is reduced 
should help to speed up the simulations.
Based on the BFM, the bond angles and bond lengths of polymers 
can be treated as dynamic degrees of freedom
depending on temperature. 
Thus, the static structure of a polymer model on the coarse grained 
level could be tuned, when one introduces bond length and bond angle 
potentials, to mimic the structure of a chemically realistic model 
of a polymer which contains covalent chemical bonds, whose orientation 
is controlled by both bond angle and torsional potentials.
In this paper we did not discuss the details of this mapping procedure 
yet, but we hope that our work will be a useful input for this problem.
However,
it will also be interesting and important to understand the distributions of bond 
lengths and torsional angles.

We hope that the present work will contribute to a
better understanding of using the lattice models for studying 
complex polymer systems and for the development of a multi-scale
coarse-graining approach based on the lattice models.

\section{Acknowledgments}

  I am indebted to K. Binder and K. Kremer for stimulating discussions.
I thank the Max Planck Institute for Polymer Research for the
hospitality while this research was carried out.
I also thank the ZDV Data Center at Johannes Gutenberg
University of Mainz for the use of the Mogon-Clusters and
the Rechenzentrum Garching (RZG), the supercomputer center of
the Max Planck Society, for the use of their computers.

\end{document}